\def\twocol{2}
\def\onecol{1}
\def\colopt{\onecol}
\ifnum\colopt=\twocol
 \documentstyle[floats,twocolumn,eqsecnum,aps,psfig,amsfonts]{revtex}
\else
 \documentstyle[floats,preprint,eqsecnum,aps,psfig,amsfonts]{revtex}
\fi
\def\cd{c^\dagger}

\def\tp{{t'}}

\def\abs#1{\left\vert #1 \right\vert}

\def\nc{n}%_c}
\def\beq{\begin{equation}}
\def\eeq{\end{equation}}
\def\bea{\begin{eqnarray}}
\def\eea{\end{eqnarray}}
\def\de{\delta}
\def\D{\Delta}
\def\la{\lambda}
\def\up{\uparrow}
\def\dn{\downarrow}

\def\nn{\nonumber}
\def\ee{{\rm e}}
\def\ie{{\it i.e.}}
\setbox122 = \hbox{1}
\def\id{\rlap{1}\rlap{\kern 1pt \vbox{\hrule width 4pt depth 0 pt}}
        \rlap{\kern 4 pt \hbox{\vrule height \ht122 depth 0 pt}}
           \hskip\wd122}
\def\Zed{\Bbb Z}

\begin{document}
\tolerance 50000
\tightenlines
\preprint{
\begin{minipage}[t]{1.8in}
\rightline{cond-mat/0008333} \rightline{La Plata-Th 00/06,
LPENS-Th 10/00, SISSA 78/2000/EP} \rightline{} \rightline{}
\end{minipage}
}

\draft

\ifnum\colopt=\twocol
 \twocolumn[\hsize\textwidth\columnwidth\hsize\csname @twocolumnfalse\endcsname
\fi

\title{
Emergence of Irrationality: Magnetization Plateaux in Modulated
Hubbard Chains
}
\author{
D.\ C.\ Cabra$^{1}$, A.\ De Martino$^{2}$\cite{PAADM}, 
A.\ Honecker$^{3}$, P.\ Pujol$^{2}$ and P.\ Simon$^{4}$\cite{PAPS}}
\address{
$^{1}$Departamento de F\'{\i}sica, Universidad Nacional de la Plata,
	C.C.\ 67, (1900) La Plata, Argentina;\\
Facultad de Ingenier\'\i a, Universidad Nacional de Lomas de Zamora,
	Cno. de Cintura y {Juan XXIII}, 
	(1832) Lomas de Zamora, Argentina.\\
$^{2}$Laboratoire de Physique\cite{URA}, 
	Groupe de Physique Th\'eorique, 
	ENS Lyon,\\ 
	46 All\'ee d'Italie, 69364 Lyon C\'edex 07, France.\\
$^{3}$Institut f\"ur Theoretische Physik, TU Braunschweig,
	Mendelssohnstr.\ 3, 38106 Braunschweig, Germany.    \\
$^{4}$International School for Advanced Studies, 
	Via Beirut 2-4, 34014 Trieste, Italy.
}
\date{August 23, 2000; revised October 17, 2000}
\maketitle
\begin{abstract}
\begin{center}
\parbox{14cm}
{Hubbard chains with periodically modulated coupling constants
in a magnetic field exhibit gaps at zero temperature in their 
magnetic and charge excitations in a variety of situations. 
In addition to fully gapped situations (plateau in the magnetization 
curve {\it and} charge gap), we have shown 
[Phys.\ Lett.\ {\bf A268}, 418 (2000)]
that plateaux also appear in the presence of massless modes, leading to 
a plateau with a magnetization $m$ whose value depends continuously
on the filling $\nc$. Here we detail and extend the arguments leading
to such doping-dependent magnetization plateaux. First we analyze the
low-lying excitations using Abelian bosonization. We compute
the susceptibility and show that due to the constraint
of fixed $\nc$, it vanishes at low temperatures (thus leading to a
magnetization plateau) even in the presence of one massless mode.
Next we study correlation functions and show that one component of the
superconducting order parameter develops quasi-long-range order on a
doping-dependent magnetization plateau.
We then use perturbation theory in the on-site repulsion $U$ to
compute the width of these plateaux up to first order in $U$.
Finally, we compute groundstate phase
diagrams and correlation functions by Lanczos diagonalization of
finite clusters, confirming the presence of doping-dependent plateaux
and their special properties.}
\end{center}
\end{abstract}

\pacs{
\hspace{-0.8cm}
}
\ifnum\colopt=\twocol
 \vskip1pc]
\fi

\section{Introduction}

Strongly correlated electron systems
in low dimensions are presently a subject of intense research.
In particular, the magnetism of such systems has revealed
very interesting properties and it is by now well established that
spin chains and  spin ladders present plateaux in their magnetization curves.
It has been shown theoretically that plateaux occur in general at
{\it rational} fractions of the saturation magnetization.
The position of these plateaux is subject to a quantization condition
that involves the volume of a translationally invariant unit cell (see
{\it e.g.}\ \cite{HiOk,AOY,Totsuka,weSpin,FGKMW,poly,YaSa}). From the
experimental side, different materials have been found which exhibit
plateaux. One example is a dimerized spin-1 chain \cite{S1dim}
which exhibits a plateau at half the saturation magnetization as predicted
in \cite{Totsuka}. A good candidate for a plateau at one third of the
saturation magnetization is Cu$_3$Cl$_6$(H$_2$O)$_2\cdot$2H$_8$C$_4$SO$_2$
\cite{Ishii}, though it is not yet fully clear whether the proposed
frustrated trimer chain model is really appropriate or which parameters
should be used \cite{fTri}.

The most striking examples of plateaux have so far
been observed experimentally in the materials
SrCu$_2$(BO$_3$)$_2$ \cite{srcu2bo32} and NH$_4$CuCl$_3$ \cite{STKT}
and both constitute again challenges for theory.
There is generally agreement that SrCu$_2$(BO$_3$)$_2$ is a predominately
two-dimensional material and how it should be modeled theoretically.
Though some progress has been made in
understanding the origin of some of the observed plateaux \cite{2Dtheory},
it remains a difficult problem to compute the complete magnetization
process within this model.
On the other hand, the high-temperature crystal structure of NH$_4$CuCl$_3$
suggests a one-dimensional model, but nevertheless it remains
unclear what really is the appropriate theoretical model for this
compound.

Materials with a ladder structure (see {\it e.g.}\ \cite{DR})
are further good candidates for exhibiting magnetization plateaux.
However, since the copper-oxide related materials are strongly coupled,
plateaux with non-zero magnetization are predicted in a magnetic field
range which causes difficulties
with the present experimental tools. A mechanism yielding plateaux
at lower values of magnetic fields would therefore be very attractive.
As we have shown in the case of modulated Hubbard chains
\cite{letter}, doping may actually provide such a mechanism since
it allows a continuous variation of the plateau magnetization $m$
with the filling $\nc$ -- extending in this particular case also into
the low-field region. Doping-dependent magnetization plateaux
have recently been also theoretically studied in a different system,
namely an integrable spin-$S$ generalization of the $t-J$ chain doped
with $(S-1/2)$ carriers \cite{FrSo} where, however, the appearance
of plateaux is restricted to large magnetization values.
Another example of such a situation occurs in the
one-dimensional Kondo lattice model \cite{KLM}
where unpaired spins behave ferromagnetically thus giving rise to
a spontaneous magnetization of a value controlled by doping.

Here we detail and extend our previous study \cite{letter} of
the effect of a magnetic field and a periodic modulation ($p$-merization)
of the hopping amplitude or the on-site energy on
a {\it doped} one-band Hubbard chain whose
Hamiltonian is given by
\bea
H &=& - \sum_{x,\alpha} t(x) \ (c^{\dagger}_{x+1,\alpha} c_{x,\alpha}
+ H.c.) + U \sum_{x=1}^L c^{\dagger}_{x,\uparrow} c_{x,\uparrow}
c^{\dagger}_{x,\downarrow} c_{x,\downarrow}
\nn \\
&& + \sum_{x,\alpha} \mu (x) \ c^{\dagger}_{x,\alpha} c_{x,\alpha}
- {h \over 2} \sum_{x=1}^L ( c^{\dagger}_{x,\uparrow} c_{x,\uparrow} -
c^{\dagger}_{x,\downarrow} c_{x,\downarrow} ) ~.
\label{pHub}
\eea
Here $c^{\dagger}_{x,\alpha}$ and $c_{x,\alpha}$ are electron creation and
annihilation operators at site $x$, $\alpha = \up,\dn$ the two spin
orientations and $h$ the external magnetic field. The hopping amplitude
$t(x)$ and the chemical potential $\mu (x)$ are taken as periodic in the
variable $x$ with period $p$.

The one-dimensional Hubbard model with dimerized coupling constants
($p=2$) is realized in a number of real compounds like the
organic (super)conductors \cite{orgS} and the ferroelectric perovskites
\cite{EIT}. While some materials in the former class come at quarter
filling, one frequently finds also realizations of the half-filled Hubbard
model. In this case, the model (\ref{pHub}) is in the same universality
class as a modulated spin-1/2 Heisenberg chain. Realizations of the
latter exist also at periods $p > 2$: Some examples of trimerized
chains ($p=3$) have been studied in \cite{BWRZZHD,AAIAG,Ishii}.

A technical motivation for resorting to the one-dimensional Hubbard model
is that the uniform chain is exactly solvable by Bethe Ansatz (BA) for
arbitrary values of the on-site repulsion $U$, filling and magnetic field
\cite{LW}. The exact solution can then be used to construct a low energy
bosonized effective field theory \cite{FK,EF,PS} which can in turn be used
to study perturbations of this model (see {\it e.g.}\ \cite{GNT}).
Here we first review some aspects of the bosonization
description of the Hubbard chain \cite{Schulz,Voit} and extend it
for the case of a finite magnetic field $h \ne 0$.

Focusing on the case of constant chemical potential $\mu(x) = \mu$,
we have shown in \cite{letter} that magnetization plateaux can appear
for the model (\ref{pHub}) if the density of particles
$\nc$ and magnetization $m$ \cite{conv} satisfy
\beq
{p \over 2} \left(\nc \pm m \right) \in \Zed  ~.
\label{cond}
\eeq
These conditions are commensurability conditions for the up electrons
$n_\up = (\nc + m)/2$ and down electrons $n_\dn = (\nc - m)/2$,
respectively.
More precisely, if both conditions are simultaneously satisfied,
the system has both charge and spin gaps. On the other hand, if
only one of these conditions is fulfilled, the filling has to be kept
fixed in order to have a magnetization plateau. 
%%%%%%%%%%%%%%%%%%%%%%%%%%%%%%%%%%%%%%%%%%%%%%%%%%%%%%%
A simple explanation of the conditions (\ref{cond}) can
be given \cite{letter} in the non-interacting limit ($U=0$). 
Then, the Hamiltonian (\ref{pHub}) can be 
easily diagonalized and is found to have $p$ bands 
$\varepsilon^\lambda(k)$ (see section \ref{smallU} for more details).
The magnetic field breaks the symmetry between up- and down-spin 
electrons by shifting their chemical potentials by opposite amounts.
It is then possible that one chemical potential (say for the up 
electrons) lies in one of the $p-1$ band gaps while the other 
(for the down electrons) is in the middle of a band. 
This situation leads to a doping-dependent plateau, if one 
imposes the constraint of fixed filling $\nc$ (and only in this case). 
Then the magnetization can be increased only by moving
an electron from the down-spin band into the up-spin band which
requires a finite energy or equivalently a finite change of magnetic field,
leading to a plateau.
However, since the filling of the down-spin electrons remains 
adjustable, one obtains a doping-dependent value of the
magnetization at the plateau.
%%%%%%%%%%%%%%%%%%%%%%%%%%%%%%%%%%%%%%%%%%%%%%%%%%%%%%%

Finally, we have also shown \cite{letter} that a charge
gap opens if the combination $p \nc \in \Zed$ of the two conditions
(\ref{cond}) is satisfied. The latter case generalizes the well known
charge gap at half filling ($\nc = 1$) as well as the charge gap
at quarter filling in the dimerized Hubbard chain ($\nc = 1/2$, $p=2$)
\cite{PM,NO99,DL99,NTO99}.

The plan of this paper is as follows: In subsection \ref{secBos}
we briefly review the bosonization approach for the Hubbard chain
for arbitrary filling and on-site Coulomb repulsion $U$ in the
presence of an external magnetic field (our conventions are
summarized in appendix \ref{convention} and details on the
bosonization approach in appendix \ref{boso}) \cite{FK,PS}. In
subsection \ref{secPoly} we then use this bosonization scheme to
study the effect of a modulation of the hopping amplitudes and the
on-site energy $\mu(x)$ and find the conditions under which a
plateau is present. The appearance of plateaux for irrational
values of the magnetization and superconducting correlations are
analyzed in subsections \ref{secPart} and \ref{secSuper},
respectively. In section \ref{smallU} we study
the limit of small $U$ perturbatively and show that the doping-dependent
plateaux are also present there.
Then we study the groundstate phase diagram (subsection \ref{secLancGS}) and
correlation functions (subsection \ref{secLancCOR}) numerically on finite size
systems by means of Lanczos diagonalization.
Finally, we summarize our results in section \ref{secConcl},
discuss some experimental settings where the features presented in this
paper could be observed and point out open routes for further
research.

\section{Bosonization approach}
\label{secAbbos}

\subsection{Field theory description of the Hubbard chain in a
magnetic field}

\label{secBos}

In this section, we summarize the analysis of the Hubbard
model in a magnetic field using Abelian bosonization.
For further details see Ref.\ \cite{PS}.
The lattice Hamiltonian is the standard one, \ie\ (\ref{pHub})
with constant $t(x) = t$, $\mu(x) = \mu$:
\bea
H &=& - t \sum_{x,\alpha} (c^{\dagger}_{x+1,\alpha} c_{x,\alpha}
+ H.c.) + U \sum_x c^{\dagger}_{x,\uparrow} c_{x,\uparrow}
c^{\dagger}_{x,\downarrow} c_{x,\downarrow}
\nn \\
&& + \mu \sum_x ( c^{\dagger}_{x,\uparrow} c_{x,\uparrow} +
c^{\dagger}_{x,\downarrow} c_{x,\downarrow} )
- {h \over 2} \sum_x ( c^{\dagger}_{x,\uparrow} c_{x,\uparrow}
- c^{\dagger}_{x,\downarrow} c_{x,\downarrow} )\,\, .
\label{HubHam}
\eea
This model has been exactly solved by BA
already in 1968 \cite{LW} but it took until 1990
for the correlation functions to be computed by combining BA results
with Conformal Field Theory (CFT) techniques \cite{FK}.
Spin-charge separation is a well known feature of the Hubbard chain at zero
magnetic field. Interestingly, it is no longer spin and charge degrees of
freedom that are separated if an external magnetic field is switched on
\cite{FK}.
Nevertheless it has been shown that in the presence of a magnetic field,
the spectrum of low energy excitations can be described by a semi-direct
product of two CFT's with central charges $c=1$ \cite{FK}.
This in turn implies that the model is still
in the universality class of the Tomonaga-Luttinger (TL) liquid and
therefore allows for a bosonization treatment.

In order to proceed, we write the fermion operator as
\bea
c_{x,\alpha} \rightarrow
\psi_{\alpha}(x) & \sim &
\ee^{i k_{F,\alpha} x}~\psi_{L,\alpha}(x)
+ \ee^{-i k_{F,\alpha} x}~\psi_{R,\alpha}(x)
~ + \ldots \\
& = & \ee^{i k_{F,\alpha} x}~\ee^{-i \sqrt{4 \pi} \phi_{L,\alpha}(x)}
+ \ee^{-i k_{F,\alpha} x}~\ee^{i \sqrt{4 \pi} \phi_{R,\alpha}(x)}
~ + \ldots ~,
\label{cfer}
\eea
where $k_{F,\alpha}$ are the Fermi momenta for up and down
spin electrons and $\phi_{R,L,\alpha}$ are the chiral components of two bosonic
fields, introduced as usual in order to bosonize the spin up and
down chiral fermion operators $\psi_{R,L,\alpha}$.
(Our conventions are settled in appendix \ref{convention}).
The dots stand for higher order terms some of which are
written explicitly in appendix \ref{boso}. They take into
account the corrections arising from the curvature of the dispersion relation
due to the Coulomb interaction. For
non-zero Hubbard repulsion $U$ and magnetic field $h$, the low
energy effective Hamiltonian corresponding to (\ref{HubHam}) written in
terms of the bosonic fields $\phi_\uparrow$ and $\phi_\downarrow$
has a complicated form, mixing up and down degrees of freedom.
The crucial step to obtain a simpler bosonized Hamiltonian is to
consider the Hamiltonian of a generalized (two component) TL model
and identify the excitations of the latter with the exact BA ones for the
model (\ref{HubHam}), providing in this way a {\it non-perturbative} bosonic
representation of the low energy sector of the full Hamiltonian (\ref{HubHam}).
This program has been carried out in Ref.\ \cite{PS} and we just quote
here the final result.
The fixed point (\ie\ neglecting all irrelevant terms) bosonized Hamiltonian
reads
\beq
H=\int dx\left[\frac{u_c}{2}
\partial_x \vec{\Phi}^t {\cal A}_c \, \partial_x\vec{\Phi}+
\frac{u_s}{2} \partial_x \vec{\Phi}^t {\cal A}_s \, \partial_x\vec{\Phi}\right] ~,
\eeq
where $\vec{\Phi}^t=(\phi_{R,\up},\phi_{L,\up},\phi_{R,\dn},\phi_{L,\dn})$.
The matrices ${\cal A}_{c,s}$ have the following form:
\beq
{\cal A}_{c,s}=\left(\begin{array}{cccc}
a_{c,s}+b_{c,s}&a_{c,s}-b_{c,s}&c_{c,s}+d_{c,s}&c_{c,s}-d_{c,s}\\
a_{c,s}-b_{c,s}&a_{c,s}+b_{c,s}&c_{c,s}-d_{c,s}&c_{c,s}+d_{c,s}\\
c_{c,s}+d_{c,s}&c_{c,s}-d_{c,s}&e_{c,s}+f_{c,s}&e_{c,s}-f_{c,s}\\
c_{c,s}-d_{c,s}&c_{c,s}+d_{c,s}&e_{c,s}-f_{c,s}&e_{c,s}+f_{c,s}\\
\end{array}\right) ~,
\eeq
where
\bea
&&\left\{
\begin{array}{ccc}
a_c=(Z_{cc}^{-1})^2 &b_c=(Z_{cc}-Z_{sc})^2 & c_c=Z_{cc}^{-1}(Z_{cc}^{-1}+Z_{cs}^{-1})\\
d_c=Z_{sc}(Z_{cc}-Z_{sc})&e_c=(Z_{cc}^{-1}+Z_{cs}^{-1})^2& f_c=Z_{cs}^2\\
\end{array}\right. \\
&&\left\{
\begin{array}{ccc}
a_s=(Z_{sc}^{-1})^2 &b_s=(Z_{cs}-Z_{ss})^2 & c_s=Z_{sc}^{-1}(Z_{ss}^{-1}+Z_{sc}^{-1})\\
d_s=Z_{ss}(Z_{cs}-Z_{ss})&e_s=(Z_{ss}^{-1}+Z_{sc}^{-1})^2& f_s=Z_{ss}^2\\
\end{array}\right.
\eea
In these expressions $Z_{ij}$ (resp. $Z^{-1}_{ij}$), $i,j=c,s$, are the
entries of the dressed charge matrix $Z$ (resp. its inverse $Z^{-1}$) taken
at the Fermi points
\beq
Z=\left(\begin{array}{cc}Z_{cc}&Z_{cs}\\Z_{sc}&Z_{ss}\\\end{array}\right) ~.
\eeq
These matrix elements are solutions of a set of coupled integral equations
obtained from the BA \cite{FK} and depend on the coupling $U$, the chemical
potential $\mu$ and the magnetic field $h$. They can be in turn related to
physical thermodynamic quantities \cite{FK}.

Substituting for the bosonic fields
\beq
\left( \begin{array}{c}
\phi_c \\
\phi_s \\
\end{array} \right)
= {1 \over \det Z }
\left( \begin{array}{cc}
Z_{ss} & Z_{ss} - Z_{cs} \\
Z_{sc} & Z_{sc} - Z_{cc} \\
\end{array} \right)
\left( \begin{array}{c}
\phi_\uparrow \\
\phi_\downarrow \\
\end{array} \right) ~,
\label{change1}
\eeq
and for their dual fields
\beq
\left( \begin{array}{c}
\theta_c \\
\theta_s \\
\end{array} \right)
=
\left( \begin{array}{cc}
Z_{cc}-Z_{sc} & Z_{sc} \\
Z_{ss}-Z_{cs} & - Z_{ss} \\
\end{array} \right)
\left( \begin{array}{c}
\theta_\uparrow \\
\theta_\downarrow \\
\end{array} \right) ~,
\label{change2}
\eeq
the Hamiltonian takes the form
\beq
\sum_{i= c,s} {u_i \over 2} \int dx ~
\left[ \left( \partial_x \phi_i \right)^2 +
\left( \partial_x \theta_i \right)^2 \right] ~,
\label{diagBos}
\eeq
where $ \phi = \phi_R + \phi_L$ and $\theta
= \phi_R - \phi_L$.

At zero magnetic field, the matrix $Z$ reduces to
\beq
Z(h=0) =
\left( \begin{array}{cc}
\xi & 0 \\
\xi /2 & 1/ \sqrt{2} \\
\end{array} \right) ~,
\label{dc0}
\eeq
with $\xi=\xi(\mu,U)$. In this case we recover the well known
expressions for the charge and spin fields
\beq
\phi_c = {1 \over \xi} \left( \phi_\uparrow + \phi_\downarrow \right) ~, ~~
\phi_s = {1 \over \sqrt{2}}
\left( \phi_\uparrow - \phi_\downarrow \right) ~,
\label{chargespin}
\eeq
where the compactification radius of the spin field (\ie\
the parameter which indicates the period of $\phi_s$, $\phi_s =
\phi_s +2\pi R_s$, $R_s = 1/\sqrt {2\pi}$)\cite{Comprad}
is fixed by the $SU(2)$
symmetry of the spin sector. The radius for the charge field, on
the other hand, depends on the chemical potential $\mu$ and the
Coulomb coupling $U$. Furthermore, for $h=0$ the charge and spin degrees
of freedom are completely decoupled.

It should be noted that for $m \ne 0$, the fields arising in the diagonalized
form of the bosonic Hamiltonian (\ref{diagBos}) are no longer the charge and
spin fields even though they have been labeled by `$c$' and `$s$'. For example,
the charge field is in general given by $\phi_\up + \phi_\dn =
Z_{cc} \phi_c - Z_{cs} \phi_s$.

For generic values of the parameters of the model (\ref{HubHam}), we can
now write down for example the bosonized expression for the charge
density operator:
\bea
\rho (x) & = & \psi^{\dagger}_\uparrow \psi_\uparrow (x) +
\psi^{\dagger}_\downarrow \psi_\downarrow (x)
\nn \\
&=& {1 \over \sqrt{\pi}} \partial_x \left( Z_{cc} \phi_c - Z_{cs} \phi_s \right)
+ a_{\rho} ~\sin [ k_+ x - \sqrt{\pi} \left( Z_{cc}\phi_c - Z_{cs} \phi_s \right) ]
\nn\\
&&\times \cos [ k_- x - \sqrt{\pi} \left( (Z_{cc}-2Z_{sc})\phi_c - (Z_{cs}-2Z_{ss})\phi_s \right)]
\nn\\
&&~+b_{\rho} \sin(2 k_+ x
-  \sqrt{4\pi} ( Z_{cc} \phi_c - Z_{cs} \phi_s )) ~, \label{density}
\eea
where $a_\rho, b_\rho$
are non-universal constants,
whose numerical values are known only in special cases.
Details on how such expressions are obtained are given in
appendix \ref{boso}. Formulae of the type (\ref{density})
are our fundamental bosonization rules.

\subsection{Space dependent modulations}
\label{secPoly}

In the present subsection we study two different perturbations
of the Hubbard chain (\ref{HubHam}), which consist in space dependent
modulations of certain parameters. In particular we shall consider a
space dependent modulation of the hopping amplitude $t(x)$ and
of the on-site energy $\mu(x)$.

\subsubsection{Modulated hopping amplitude}\label{hopping}

In this case, the Hamiltonian reads as in eq.\ (\ref{pHub}) with
$\mu(x)= const$ and $t(x) = t$ if $x \neq lp$ and $t(lp) = t' = t + \delta$,
with $p,l$ integers and $p$ fixed. This is equivalent to the uniform Hubbard
Hamiltonian (\ref{HubHam}) perturbed by the term
\beq
H_{pert} = - \de \sum_{x'=lp,\alpha}
 (c^{\dagger}_{x',\alpha} c_{x'+1,\alpha} + H.c.) ~.
\label{perturb}
\eeq
At half filling and for large $U$, a standard second
order perturbative computation in $1/U$ shows that the effective
Hamiltonian is given by
\beq
\tilde{H} = \sum_{x} {4 t^2(x) \over U}~ \vec{S}_x \cdot \vec{S}_{x+1} ~,
\label{JtU}
\eeq
thus leading to the $p$-merized Heisenberg chain studied in
\cite{poly}. It was predicted there that magnetization plateaux occur
when the condition ${p\over 2}(1-m) \in \Zed$ is
satisfied \cite{conv}.
We use now Abelian bosonization techniques to analyze the more general
case of the model (\ref{pHub}) in the small $\de$ (weak $p$-merization) limit.

Using the bosonization dictionary given in appendix \ref{boso}, we find
the expression for the continuum limit of the lattice perturbation
(\ref{perturb})
\bea
O_{pert} &=& ~
 \lambda_1 \sin [k_+ /2 + p k_+ x - \sqrt{\pi} \left( Z_{cc}\phi_c -
Z_{cs} \phi_s \right) ]
\nn\\
&& \times \cos [k_- /2 + p k_- x - \sqrt{\pi} \left( (Z_{cc}-2Z_{sc})\phi_c -
(Z_{cs}-2Z_{ss})\phi_s \right)]
\nn\\
&& + \lambda_2 \sin[ k_+ + 2 p k_+ x
-  \sqrt{4\pi} ( Z_{cc} \phi_c - Z_{cs} \phi_s ) ] ~,
\label{pert}
\eea
where $\la_1,\la_2\propto \delta$ and the Fermi momenta are
$k_+=k_{F,\up}+k_{F,\dn}=\pi n$, $k_-=k_{F,\up}-k_{F,\dn}=\pi m$,
where $n$ is the filling and $m$ is the magnetization.
The presence of a factor $p$ in the oscillating part
will play an important role in the following.

The operator
\beq
\lambda_3 \cos [k_- + 2p k_- x - 2\sqrt{\pi} \left( (Z_{cc}-2Z_{sc})
\phi_c - (Z_{cs}-2Z_{ss})\phi_s \right)] ~,
\label{pert2}
\eeq
with $\la_3 \propto \de^2$ is radiatively generated from
the first term in (\ref{pert}) and must therefore be included as well.

In the case of zero magnetic field the dressed charge matrix is
given by (\ref{dc0}) and we have then a neat separation between charge and
spin fields. The most relevant perturbation takes the form
\beq
O_{pert} =
\lambda_1 \sin [{\pi n\over 2} + p n \pi x - \sqrt{\pi}\xi \phi_c]
\cos [\sqrt{2\pi}\phi_s ] + \lambda_2 \sin[ \pi n + 2 p n \pi x
-  \sqrt{4\pi} \xi \phi_c] ~.
\label{perth0}
\eeq
The marginal operator associated with $\lambda_3$ contains only the
spin field, its dimension (fixed by the $SU(2)$ symmetry) is 2 and
it is marginally irrelevant. A term like this is already present
in the original model and is also marginally irrelevant. Hence,
for $\delta$ small enough, this term can be absorbed in the
original marginally irrelevant perturbation term without changing
its relevance character.

The $\lambda_2$ term affects only the charge degrees of freedom and
its dimension runs from $1$, for $U \rightarrow \infty$ to $2$,
for $U=0$, being then always relevant for the cases of interest.
We can therefore conclude that the charge field is massive
whenever this operator is commensurate, which in turn happens if
the condition $p n \in \Zed$ is satisfied.

If this happens, we can integrate out the massive charge degrees of
freedom which leaves us with an effective theory for the spin
degrees of freedom. This effective theory is massless except
when the operator associated with $\lambda_1$ becomes
also commensurate, \ie\ if the condition ${p n \over 2} \in \Zed$
is satisfied. In that case, we have also a spin gap in the system.

These considerations are easily generalized to the case
of non-zero magnetization as long as the condition $p n \in \Zed$ is satisfied.
In this case, the $\lambda_2$ term in (\ref{pert}) is always commensurate.
Since it contains only the proper charge field $\phi_\up + \phi_\dn$, a charge
gap opens for all values of $m$ at these commensurate values of the filling.
The condition $p n \in \Zed$ is also satisfied when the two conditions
(\ref{cond}) are simultaneously satisfied. In this case, also the
$\lambda_1$ term in (\ref{perth0}) becomes commensurate thus leading
also to a spin gap.

In particular for $p=2,3$, we predict the following fully gapped
situations:
\begin{itemize}
\item[$p=2$:]
Half filling ($n =1$): gap for the charge, and plateau for $m=0$. \\
Quarter filling ($n =1/2$) (and also $n=3/2$): gap for the charge
\cite{PM,NO99,DL99,NTO99}, and plateau for $m= \pm 1/2$.
\item[$p=3$:]
$n =1$, $n =1/3$ and $n =5/3$: gap for the charge, and
plateau for $m= \pm 1/3$. \\
$n =2/3$ and $n =4/3$: gap for the charge, and plateau for
$m=\pm 2/3$, $0$.
\end{itemize}

The final case where only one of the conditions (\ref{cond}) holds is more
complicated since then the charge and spin degrees of freedom can
no longer be separated. We therefore postpone discussion of this
case a little.

\subsubsection{Modulated on-site energy} \label{onsite}

Now we consider the Hubbard chain (\ref{pHub}) with a uniform hopping amplitude
$ t(x) = t $ but a periodic modulation of the chemical potential
$\mu (x) = \mu$ if $x \neq lp$ and $\mu (lp) = \mu + \delta\mu$,
with $p,l$ integers, $p$ fixed. This is equivalent to the uniform chain
(\ref{HubHam}) plus an on-site energy term which reads
\beq
H'_{pert} = \delta \mu \sum_{x'=lp,\alpha} c^{\dagger}_{x,\alpha} c_{x,\alpha} ~.
\label{pHub1}
\eeq
The case $p=2$, $h=0$ has been studied in detail in \cite{FGN}.

In the continuum limit the perturbing operator (\ref{pHub1}) becomes:
\bea
O'_{pert} & = &
\lambda_1 \sin [p k_+ x - \sqrt{\pi} \left( Z_{cc}\phi_c -
Z_{cs} \phi_s \right) ]
\cos [ p k_- x - \sqrt{\pi} \left( (Z_{cc}-2Z_{sc})\phi_c -
(Z_{cs}-2Z_{ss})\phi_s \right)]
\nn\\
&&+\lambda_2 \sin( 2 p k_+ x
-  \sqrt{4\pi} ( Z_{cc} \phi_c - Z_{cs} \phi_s ) ) ~,
\label{pert'}
\eea
which radiatively generates a term of the form
\beq
\lambda_3 \cos [2p k_- x - 2\sqrt{\pi} \left( (Z_{cc}-2Z_{sc})
\phi_c - (Z_{cs}-2Z_{ss})\phi_s \right)] ~.
\eeq
The only difference with respect to the previous case,
eqs.\ (\ref{pert}), (\ref{pert2}), is a phase in each of the sines
or cosines, which however plays a role only at half filling and
zero magnetic field (see also \cite{FGN}). Apart from this
particular case, the conclusions remain the same as in the
previous case.

\subsection{Partial gap: irrational plateaux}
\label{secPart}

We have shown in the previous subsection that, when both commensurability
conditions (\ref{cond}) are satisfied, the spectrum of the model (\ref{pHub})
is fully gapped. It remains to understand what happens if only one of these
conditions is satisfied. In this case, apparently one degree of
freedom remains massless. We will show that the system nevertheless
exhibits a gap for magnetic excitations ($\ie$ excitations changing
the value of the magnetization), provided the total charge
(\ie\ the filling $\nc$) is kept fixed.

For the sake of simplicity we will restrict ourselves to the dimerized
chain ($ p=2 $) in this subsection though the argument can be generalized
easily to $p > 2$. Suppose that
\beq
 \frac{p}{2} (n  - m) = n - m = 2 n_{\dn} \in \Zed ~,
\label{commens}
\eeq
but that $n+m$ is not an integer, \ie\
the commensurability condition is fulfilled only for down electrons.
Assume first that there is no interaction between up and down
electrons ($U=0$). Then we are led to analyze the excitation spectrum 
of the following Hamiltonian in a system of length $L$:
\beq
H = \int_{0}^{L} dx ~ {v_\uparrow \over 2} ~ \left[ \left( \partial_x
\phi_\uparrow \right)^2 +
\left( \partial_x \theta_\uparrow \right)^2 \right] +
~ {v_\downarrow \over 2} ~
\left[ \left( \partial_x \phi_\downarrow \right)^2 +
\left( \partial_x \theta_\downarrow \right)^2 \right] +
\lambda \cos( 2 \sqrt{\pi} \phi_\downarrow ) ~,
\label{ham}
\eeq
with total  magnetization
\beq
M = \int _{0}^{L} dx ~ \frac{1}{\sqrt{\pi}} \partial_x ( \phi_\uparrow -
\phi_\downarrow ) = \frac{1}{\sqrt{\pi}} ( \phi_\uparrow - \phi_\downarrow )
|_0^L ~.
\label{M}
\eeq
Motivated by experimental realizations of
Hubbard systems, where typically doping is fixed, we also
impose the constraint that the total particle number
is fixed:
\beq
N = \int dx ~ \frac{1}{\sqrt{\pi}} \partial_x
( \phi_\uparrow + \phi_\downarrow ) = \frac{1}{\sqrt{\pi}}
( \phi_\uparrow + \phi_\downarrow ) |_0^L ~.
\label{Nc}
\eeq
 From eqs.\ (\ref{M}) and (\ref{Nc}) we see that the fields
$\phi_{\uparrow,\downarrow}$ satisfy the following boundary
conditions:
\bea
2 \phi_\uparrow |_0^L = \sqrt{\pi} ( N + M ) ~, \label{upbc} \\
2 \phi_\downarrow |_0^L = \sqrt{\pi} ( N - M ) ~. \label{downbc}
\eea
Notice, furthermore, that the fields $\phi_{\uparrow,\downarrow}$
are compactified, \ie\ they satisfy the periodicity condition
\beq
\phi_{\uparrow,\downarrow} \rightarrow \phi_{\uparrow,\downarrow}
+ \sqrt{\pi} \Zed ~.
\eeq
Therefore, in a semiclassical picture,
the vacuum configuration for $\phi_\uparrow$ is
\beq
\phi_\uparrow (x) = \frac{\sqrt{\pi}}{2 L} ( N + M ) x + const  ~.
\eeq
On the other hand, for $\phi_\downarrow$ the vacuum configuration is a kink
\beq
\phi_\downarrow (x) = k(x) ~,
\eeq
where $k(x)$ is a configuration interpolating between two minima
of the cosine potential in the Hamiltonian (\ref{ham}) and satisfying
the boundary condition (\ref{downbc}).
%Notice that consistency requires $(N - M ) = 2 \Zed$ which generically is
%only possible if $L$ is even.
Now we change the total magnetization keeping the total number of particles
fixed. The lowest energy excitation of this type consists in reversing
the spin of a particle, which corresponds to the
change $ M \rightarrow M+2$. The new boundary conditions become then
\bea
2 \phi_\uparrow |_0^L = \sqrt{\pi} ( N + M +2 ) ~,\label{upbc2} \\
2 \phi_\downarrow |_0^L = \sqrt{\pi} ( N - M -2) ~.\label{downbc2}
\eea
The new vacuum configuration for $\phi_\uparrow$ is therefore
\beq
\phi_\uparrow (x) = \frac{\sqrt{\pi}}{2 L} ( N + M + 2 ) x + const ~,
\eeq
and it is straightforward to show that the difference
in energy with respect to the original one is linear in $1/L$.
On the contrary,  changing the configuration of the kink
requires a finite amount of energy (proportional to $\lambda$) because
the new configuration is in a different topological sector.
This corresponds to the presence of a gap in the spectrum of magnetic 
excitations, and therefore of a plateau in the magnetization curve.

To support the previous conclusion further, we analyze the
magnetic susceptibility for a chain of finite size $L$. It is given
by the integral of the correlation function of the spin density
operator ${1 \over \sqrt{\pi}} \partial_x ( \phi_\uparrow -
\phi_\downarrow )$.
Since the down sector is gapped  it does not contribute to the
zero temperature limit of the susceptibility. Let us therefore
focus on the up sector, which is apparently massless but
constrained to be in a particular topological sector. One can
easily see that determination of the susceptibility amounts to calculate
\beq
\left\langle \left(\int_0^L dx~\partial_x \phi_\uparrow \right)^2 \right\rangle
-
\left\langle \left(\int_0^L dx~\partial_x \phi_\uparrow \right) \right\rangle^2
\label{suscept} \,\, .
\eeq
For the free massless sector, the Hamiltonian in a finite size $L$
can be written in Fourier space for each topological sector as
(see for example \cite{cdm}):
\beq
H_L = \frac{v}{2} \sum_{q \neq 0}
\left[ \frac{1}{K} q^2 \phi_{-q} \phi_{q} + K q^2 \theta_{-q}\theta_{q} \right]
+ { \pi v \over 2 L} \left( \frac{1}{K} Q^2 + K J^2 \right) ~,
\eeq
where $Q$ and $J$ stand for the particle number and current
zero modes:
$$Q = {1 \over \sqrt{\pi}} \left. \phi_\up \right\vert_0^L ~, \qquad
J = {1 \over \sqrt{\pi}} \left. \theta_\up \right\vert_0^L ~,
$$
and the summation over $q$ is for oscillatory modes. If the global
constraint is not present, one has to sum over all possible values
of $Q$, \ie\ one has to compute
\beq \chi = {1 \over \beta L}
\left( {1 \over Z} \mbox{Tr} \left(\exp( - \beta H_L) Q^2 \right) -
\left({1 \over Z} \mbox{Tr} \left(\exp( - \beta H_L) Q \right)\right)^2
\right) ~.
\eeq
The local part (the oscillator modes) decouples, and if the constraint is
not imposed we obtain the standard result
\beq \chi = {1 \over \beta L} \left( {1 \over
Z} \sum_{Q} \left(\exp( - \beta { \pi v Q^2 \over 2 LK}  ) Q^2
\right) - \left({1 \over Z} \sum_{Q} \left(\exp( - \beta { \pi v
Q^2 \over 2 LK} ) Q \right)\right)^2 \right) = {K \over 2 \pi v} ~.
\eeq
If we now impose the global constraint on $N$, due to the gap in
the down sector as discussed above, all the sectors will be
exponentially suppressed, except the sector
$
Q = ( N + M )/2 $. Therefore, for small enough temperature, the
distribution $\exp( - \beta { \pi v Q^2 \over 2 LK} )$ has to be replaced
by a delta function in $Q = ( N + M )/2$, giving
\beq
\chi = {1 \over \beta L} \left( \langle Q^2 \rangle -
\langle Q \rangle^2 \right) = 0 ~.
\eeq
We find then an exotic situation in which we have simultaneously
algebraic decay of correlation functions, since the local dynamics is
massless, but zero magnetic susceptibility, due to the global constraint
imposed on the system. The only somewhat similar situations we are
aware of include plateau states of strongly frustrated spin ladders with
gapless non-magnetic excitations \cite{weSpin,Tan,nanoT} as well as the large
number of singlets inside the gap of the Heisenberg antiferromagnet
on a Kagom\'e lattice \cite{Kagome}.

\subsection{Superconducting fluctuations}
\label{secSuper}

Having found a situation with a gap which can be attributed to magnetic
excitations and another massless degree of freedom, one may wonder whether
superconducting fluctuations develop. We therefore now briefly analyze the
correlators of the superconducting order parameter. In the presence of
a magnetic field, the superconducting order parameter has four components
which read on the lattice:
\beq
\Delta_{\alpha,\beta} = c_{x+1,\alpha} c_{x,\beta} \, .
\label{defSC}
\eeq
For $h=0$, these components can be grouped in
a triplet ($t$) and a singlet ($s$). On the lattice, the corresponding $S^z = 0$
components can be chosen as
\beq
\Delta_{t,s}^{latt} = c_{x,\uparrow} c_{x+1,\downarrow}
\pm  c_{x,\downarrow} c_{x+1,\uparrow}\, .
\label{sctlatt}
\eeq
In the continuum, using eq.\ (\ref{psiRL}) this leads to the following
expression
\bea
\Delta_{t,s} &=& \ee^{-ik_-x} \psi_{R,\uparrow} \psi_{L,\downarrow}
\,( \ee^{ik_{F\downarrow}} \mp  \ee^{-ik_{F\uparrow}} )
+ \ee^{ik_-x} \psi_{L,\uparrow} \psi_{R,\downarrow}
\,( \ee^{-ik_{F\downarrow}} \mp \ee^{ik_{F\uparrow}} ) \nn\\
& + & \ee^{-ik_+x} \psi_{R,\uparrow} \psi_{R,\downarrow}
\,( \ee^{-ik_{F\downarrow}} \mp  \ee^{-ik_{F\uparrow}} )
+ \ee^{ik_+x}\psi_{L,\uparrow} \psi_{L,\downarrow}
\,( \ee^{ik_{F\downarrow}} \mp \ee^{ik_{F\uparrow}} )  ~.
\label{scgen}
\eea
In particular, for zero magnetic field
$k_{F\uparrow} = k_{F\downarrow}= k_F$ and neglecting
``$2k_F$'' terms, eqs. (\ref{scgen})
reduce to the standard ones (see for example Ref.\ \cite{GNT}):
\beq
\Delta_{t} = 2i \sin k_F ( \psi_{R,\uparrow} \psi_{L,\downarrow}
+ \psi_{R,\downarrow} \psi_{L,\uparrow} ) \, , \qquad
\Delta_{s} = 2 \cos k_F ( \psi_{R,\uparrow} \psi_{L,\downarrow}
-\psi_{R,\downarrow} \psi_{L,\uparrow}(x) \, .
\label{scStandard}
\eeq
For general $m$, one finds instead
\beq
\Delta_{t} \sim \frac{1}{\pi
a} \ee^{i\sqrt{\pi}(\theta_\uparrow + \theta_\downarrow)} \cos
(\sqrt\pi (\phi_\uparrow - \phi_\downarrow) - k_- x) \, , \quad
\Delta_{s}  \sim \frac{i}{\pi a}
\ee^{i\sqrt{\pi}(\theta_\uparrow + \theta_\downarrow)} \sin
(\sqrt\pi (\phi_\uparrow - \phi_\downarrow) - k_- x) \, .
\label{scMne0}
\eeq
These expressions show that the correlators associated to 
the order parameters $\Delta_{t}$ and $\Delta_{s}$ decay 
exponentially, even in the partially massless plateau phases. 
Indeed, since the $S^z = 0$ components of $\Delta_{\alpha,\beta}$ 
are products of up and down degrees of freedom, 
it is sufficient that one of them is gapped in order to lead 
to an exponential decay of the composite object.

On the other hand, the diagonal components are bosonized as
\beq
\Delta_{\alpha,\alpha} \sim  2 \cos(k_{F,\alpha}) \ee^{i \sqrt{4 \pi} \theta_\alpha}
+ \ee^{-i k_{F,\alpha} (1+2 x)}
\ee^{i \sqrt{4 \pi} (\phi_\alpha + \theta_\alpha)}
+ \ee^{i k_{F,\alpha} (1+2 x)}
\ee^{i \sqrt{4 \pi} (-\phi_\alpha + \theta_\alpha)} + ... \, ,
\label{bosDaa}
\eeq
where the dots include terms which mix $\phi_\up$ with $\phi_\dn$.
It is then clear that on a doping-dependent plateau, where only one
of the fields is gapful, only one of the correlators
$\langle \Delta_{\alpha,\alpha}^\dagger \Delta_{\alpha,\alpha} \rangle$
decays exponentially,
but the other exhibits {\it algebraic} behavior. In fact, all 
fields involving only the gapless spin component decay
algebraically. In particular, also the two-point correlator of
$c_{x,\alpha}$ decays algebraically if $\Delta_{\alpha,\alpha}$
exhibits quasi-long-range order.
The algebraic decay of the latter should therefore not be taken as a
sign for superconductivity, but is interesting nevertheless.
%Although it remains to be
%clarified to which extent this should be taken as a sign for true
%superconductivity, it is certainly interesting that one of the components
%of the superconducting order parameter exhibits quasi-long-range order.

\section{Small-$U$ limit} \label{smallU}

The previous section was dedicated to the bosonization
approach to the $p$-merized Hubbard chain in the small $p$-merization
limit. In the present section, we give a further argument for
doping dependent plateaux, valid in the low $U$ limit but at arbitrary
$p$-merization strength. For the sake of simplicity, we will concentrate
on the case of modulated hopping amplitude $t(x) = t'$ for
$x$ a multiple of $p$, otherwise $t(x) = t$ and constant $\mu(x) = \mu$,
but the arguments can be generalized easily.

First we diagonalize the Hamiltonian (\ref{pHub})
at $U=0$ by a unitary transformation
\beq
d^{\lambda}_{k,\sigma} = {1 \over \sqrt{L}} \sum_{x=1}^{L/p}
  \ee^{i k x} \sum_{j=1}^p a^{\lambda}_{k,j}
   c_{xp+j,\sigma} \, .
\label{su2}
\eeq
In order for the kinetic part of the Hamiltonian (\ref{pHub})
to take the form
\beq
H_0 = \sum_{\lambda=1}^p \sum_{\sigma} \epsilon^{\lambda}(k)
    d^{\dagger\lambda}_{k,\sigma} d^{\lambda}_{k,\sigma} \, ,
\label{su3}
\eeq
the coefficients $a^{\lambda}_{k,j}$ have to satisfy the following
eigenvalue equation:
\beq - \left(
  \matrix{0 & t & 0 & \cdots & 0 & t' \ee^{-i k} \cr
          t & 0 & t & \ddots &  & 0 \cr
          0 & t & \ddots & \ddots & \ddots & \vdots \cr
          \vdots & \ddots & \ddots & \ddots & t & 0 \cr
          0 &  & \ddots & t & 0 & t \cr
          t' \ee^{i k} & 0 & \cdots & 0 & t & 0 \cr
}\right) \left(\matrix{a^{\lambda}_{k,1} \cr
     a^{\lambda}_{k,2} \cr \vdots \cr a^{\lambda}_{k,p}}
    \right)
= \epsilon^{\lambda}(k)
\left(\matrix{a^{\lambda}_{k,1} \cr a^{\lambda}_{k,2} \cr
    \vdots \cr a^{\lambda}_{k,p}}
    \right) \, .
\label{su4}
\eeq
The resulting $p$ energy bands $\epsilon^{\lambda}(k)$ are illustrated
in Fig.\ 2 of \cite{PM} for $p=2$ and in Fig.\ 2 of \cite{letter}
for $p=3$ (note that in the latter case, the energy $\epsilon$
was plotted with the wrong sign which can be absorbed by shifting
$k \to k+\pi$).

In the sequel we first work out the simpler case $p=2$ and then
generalize to $p\geq 3$.

\subsection{Case $p=2$}

In the dimerized case, the eigenvalue equation (\ref{su4}) reduces to
\beq - \left(
  \matrix{0 & t + t' \ee^{-i k} \cr
          t + t' \ee^{i k} & 0 \cr
}\right) \left(\matrix{a^{\lambda}_{k,1} \cr a^{\lambda}_{k,2} \cr
    }\right)
= \epsilon^{\lambda}(k)
\left(\matrix{a^{\lambda}_{k,1} \cr a^{\lambda}_{k,2} \cr
    }\right) \, .
\label{su5}
\eeq
The eigenvalue problem (\ref{su5}) is solved readily, yielding
\bea
\epsilon^{\pm}(k) & = & \pm \sqrt{ t^2 + t'^2 + 2 t t'
\cos{k} } \, , \nonumber \\
a^{\pm}_1 & = & \mp \sqrt[4]{ \frac{ t + t' {\rm
e}^{-i k} }{ t + t' \ee^{i k} } } \, , \label{su6} \\
a^{\pm}_2 & = & \sqrt[4]{ \frac{ t + t' \ee^{i k} }
   { t + t' \ee^{-i k} } }
\, . \nonumber
\eea
The inverse of the transformation (\ref{su2})
is:
\bea
c_{2x+2,\sigma} & = & {1 \over \sqrt{L}}
  \sum_{k} \ee^{- i k x}
     \sqrt[4]{{t + t' \ee^{-i k} \over t + t' \ee^{i k}}}
   \left(d^{-}_{k,\sigma} + d^{+}_{k,\sigma}\right) \, , \nn \\
c_{2x+1,\sigma} & = & {1 \over \sqrt{L}}
  \sum_{k} \ee^{- i k x}
     \sqrt[4]{{t + t' \ee^{i k} \over t + t' \ee^{-i k}}}
   \left(d^{-}_{k,\sigma} - d^{+}_{k,\sigma}\right) \, .
\label{su7}
\eea
Eigenstates $| \{k^\lambda_{j,\sigma}\} \rangle$
of the free Hamiltonian $H_0$ are now  written
down by simply specifying the momenta $k^\lambda_{j,\sigma}$
occupied in the various bands. Now we treat the Coulomb repulsion
\beq
H_I = U \sum_{x=1}^L n_{x, \up} n_{x,\dn}
\label{su8}
\eeq
in first order perturbation theory.

To proceed further, we assume that none of the bands are half-filled.
Then $H_I$ has only diagonal terms (\ie\ Umklapp scattering is absent)
which are readily evaluated as (denoting
$n^{\lambda}_{k,\sigma} = d^{\dagger\lambda}_{k,\sigma} d^{\lambda}_{k,\sigma}$)
\bea
\langle \{ k^\lambda_{j,\sigma} \} | H_I | \{ k^\lambda_{j,\sigma} \} \rangle
& = & {U \over L}
\langle \{ k^\lambda_{j,\sigma} \} |
\sum_{k,k'}
(n^-_{k,\up} + n^+_{k,\up}) (n^-_{k',\dn} + n^+_{k',\dn})
| \{ k^\lambda_{j,\sigma} \} \rangle  \nn \\
& = & {U \over L} (N^-_{\up} + N^+_{\up})
         (N^-_{\dn} + N^+_{\dn}) \nn \\
& = & {U \over 4} L (n^-_{\up} + n^+_{\up})
         (n^-_{\dn} + n^+_{\dn}) \, .
\label{su9}
\eea
In the second line, we have defined the number of particles with
spin $\sigma$ in band $\lambda$ by $N^{\lambda}_{\sigma} =
n^{\lambda}_{\sigma} L /2$\cite{totnum}. The densities have been  normalized
such that $n^{\lambda}_{\sigma} =1$ for a completely filled band.

Similarly, expectation values of the number operators
$\sum_{x=1}^L n_{x,\sigma}$ give $L \left(n^{-}_{\sigma} +
n^{+}_{\sigma}\right)/2$. Putting everything together, we find the
energy of the Hamiltonian (\ref{pHub}) to first order in $U$ as
\bea
E & = & \sum_{\lambda} \sum_{k^\lambda_{j,\sigma}}
  \epsilon^\lambda(k^\lambda_{j,\sigma})
 + {U \over 4} L (n^-_{\up} + n^+_{\up})
         (n^-_{\dn} + n^+_{\dn}) \nn \\
 && +  {\mu  \over 2} L (n^-_{\up} + n^+_{\up} + n^-_{\dn} + n^+_{\dn})
 - {h \over 4} L (n^-_{\up} + n^+_{\up} - n^-_{\dn} - n^+_{\dn}) \, .
\label{su10}
\eea
Assume now that the `$-$' bands are both
partially filled. Then (\ref{su10}) specializes to
\beq
E/L = {1 \over 4 \pi} \sum_{\sigma}
\int\limits_{-n_\sigma \pi}^{n_\sigma \pi} d k \epsilon^{-}(k)
    + {U \over 4} n_\up n_\dn
    + {\mu \over 2} (n_\up+n_\dn) - {h \over 4} (n_\up-n_\dn) \, ,
\label{su11}
\eeq
where $n_{\sigma} = n^-_{\sigma}$ \cite{normden}. Setting $n = n_\up+n_\dn$ and
fixing $n_\up \approx 1$, we find from the condition that it does
not require energy to flip $\up$ to $\dn$ spins,
\beq
h_{c_1} =  \epsilon^-(\pi) - \epsilon^-((n-1)\pi)
          + U \left({n\over2}-1\right) ~.
\label{su12}
\eeq
On the other hand, if we consider the case of a completely filled
``$-,\up$'' band and partially filled ``$+,\up$'' and ``$-,\dn$'' bands,
eq.\ (\ref{su10}) specializes as follows:
\bea
E/L  & = & {1 \over 2 \pi}
 \int\limits_{(1-n^+_{\up})\pi}^{n^+_{\up} \pi} d k \epsilon^{+}(k)
+ {1 \over 2 \pi} \int\limits_{0}^{\pi} d k \epsilon^{-}(k)
+ {1 \over 2 \pi} \int\limits_{0}^{n_{\dn} \pi} d k \epsilon^{-}(k)
 + {U \over 4} (1+n^+_{\up}) n_{\dn} \nn \\
 && +  {\mu \over 2} (1+ n^+_{\up} + n_{\dn})
 - {h \over 4} (1 + n^+_{\up} - n_{\dn}) \, .
\label{su13}
\eea
Setting $n = 1+n^+_{\up}+n_\dn$ and fixing $n^+_{\up} \approx 0$, we find
in the same way as before that
\beq
h_{c_2} = \epsilon^+(\pi) -  \epsilon^-((n-1)\pi)
          + U \left({n\over2}-1\right) \, .
\label{su14}
\eeq
Using eqs.\ (\ref{su14}) and (\ref{su12}) we find that the width of
the plateau at fixed $N$ is not affected by the on-site Coulomb
repulsion to first order in $U$:
\beq
h_{c_2} - h_{c_1} = 2 \abs{t-t'} + {\cal O}(U^2)\, .
\label{su15}
\eeq
This ensures the presence of a doping-dependent plateau with $m = 1-n$
in the low $U$ limit. The absence of a first-order correction in $U$ to the
width in (\ref{su15}) can be traced to the mean-field form
(\ref{su9}) of the matrix elements of the on-site repulsion $H_I$.
This in turn is due to the fact that $\abs{a^{\pm}_i}=1$ for all
$k$ as can be seen from (\ref{su6}) and is a manifestation
of the symmetry between the lower and upper band. Both the mean-field
form of the interaction as well as the absence of a first-order
correction to the plateau width are special properties of the case
$p=2$, as will become clear in the following discussion of the case
$p \geq 3$.

\subsection{Case $p\ge 3$}

For general $p$, the diagonalization (\ref{su4}) is more complicated leading
to the absence of explicit expressions such as (\ref{su7}).
Nevertheless, we can still use unitarity of the transformation (\ref{su2})
to formally invert it
\beq
c_{xp+j,\sigma}={1\over \sqrt{L}} \sum_k\sum_{\la} \ee^{-ikx}
a_{k,j}^{*\la} d_{k,\sigma}^{\la} \, .
\label{genInv}
\eeq
First, we look at the transformation of number operators
$n_{x,\sigma} = \cd_{x,\sigma} c_{x,\sigma} \to
n^{\lambda}_{k,\sigma} = d^{\dagger\lambda}_{k,\sigma}
d^{\lambda}_{k,\sigma}$:
\beq
\sum_{x} n_{x,\sigma} = {1 \over p}
   \sum_{k,\la} \sum_{j=1}^p \abs{a^{\la}_{k,j}}^2 n^{\la}_{k,\sigma}
= \sum_{k,\la} n^{\la}_{k,\sigma} \, .
\label{nInK}
\eeq
Here, we have just used that $\sum_{j=1}^p \abs{a^{\la}_{k,j}}^2 = p$.

The diagonal terms of the interaction (\ref{su8}) can now be treated
similarly as for $p=2$. Instead of (\ref{su9}) one finds for general $p$
\bea
\langle \{ k^\lambda_{j,\sigma} \} | H_I | \{ k^\lambda_{j,\sigma} \} \rangle
& = & {U \over p L}
\langle \{ k^\lambda_{j,\sigma} \} | \sum_{j=1}^p
\sum_{k,\la} \abs{a^{\la}_{k,j}}^2 n^{\la}_{k,\up}
\sum_{k',\la'} \abs{a^{\la'}_{k',j}}^2n^{\la'}_{k',\dn}
| \{ k^\lambda_{j,\sigma} \} \rangle \nn\\
& = &  {U \over p L} \sum_{j=1}^p
\sum_{{k,\la} \atop k^\lambda_{j,\up}\ {\rm occupied}}
           \abs{a^{\la}_{k,j}}^2
\sum_{k',\la' \atop {k'}^{\lambda'}_{j,\dn}\ {\rm occupied}}
           \abs{a^{\la'}_{k',j}}^2 \, .
\eea
Next we pass to the thermodynamic limit which leads
to replacing sums by integrals. Due to (\ref{nInK}), integrals and
differentials over densities can be replaced by integrals in
$k$-space. We work at a fixed particle number $n$, which implies
\beq
0 = d n = d n_{\up} + d n_{\dn} \qquad \Rightarrow \qquad
   d n_{\dn} = - d n_{\up} \, .
\eeq
Now we  concentrate on the situation where all bands $\la \le
\la_0$ are completely occupied with up spins and those with $\la >
\la_0$ do not contain any up spins generalizing thus the reasoning
of the previous section. The band $\la_0'$ is partially occupied
with down spins, those with $\la' < \la_0'$ are completely filled
with down spins while those with $\la' > \la_0'$ do not contain
any down spins. For a partially filled band $\nu$ denote finally
the occupied states by $[k_l^\nu,k_u^\nu]$.

Then we can generalize (\ref{su11}) to first order in $U$ as
follows\cite{normden}
\beq
E/L = u + v + {\mu \over p} \left(n_\up + n_\dn\right)
    - {h \over 2 p} \left(n_\up - n_\dn\right)
\eeq
with
\beq
u = {1 \over 2 \pi p} \left\{\sum_{\la \le \la_0}
   \int_{-\pi}^{\pi} d k \epsilon^\la(k)
   + \int_{k_l^{\la_0'}}^{k_u^{\la_0'}} d k' \epsilon^{\la_0'}(k')
   + \sum_{\la' < \la_0'} \int_{-\pi}^{\pi} d k' \epsilon^{\la'}(k') \right\}
\eeq
and
\beq
v = {U \over 4 \pi^2 p^3} \left\{ \sum_{j=1}^p
  \sum_{\la \le \la_0} \int_{-\pi}^{\pi} d k \abs{a_{k,j}^\la}^2
  \left(
  \int_{k_l^{\la_0'}}^{k_u^{\la_0'}} d k' \abs{a_{k',j}^{\la_0'}}^2
   + \sum_{\la' < \la_0'} \int_{-\pi}^{\pi} d k' \abs{a_{k',j}^{\la'}}^2
  \right)
\right\} \, .
\eeq
This yields for the lower boundary of the associated plateau
\beq
h_{c_1} = \epsilon^{\la_0}({k_u^{\la_0}}) - \epsilon^{\la_0'}(k_u^{\la_0'})
+ {\cal U}(\la_0) + {\cal O}\left(U^2\right)
\label{hc1genP}
\eeq
with
\bea
{\cal U}(\nu) &=&
{U \over 2 \pi p^2} \left\{
\sum_{j=1}^p \abs{a_{k_u^\nu,j}^{\nu}}^2
  \left(
  \int_{k_l^{\la_0'}}^{k_u^{\la_0'}} d k' \abs{a_{k',j}^{\la_0'}}^2
   + \sum_{\la' < \la_0'} \int_{-\pi}^{\pi} d k' \abs{a_{k',j}^{\la'}}^2
  \right)\right.
\nn\\ && \qquad \qquad
\left.- \sum_{j=1}^p \sum_{\la \le \la_0}
\int_{-\pi}^{\pi} d k \abs{a_{k,j}^\la}^2
   \abs{a_{k_u^{\la_0'},j}^{\la_0'}}^2
\right\} \, .
\eea
For the corresponding upper boundary one finds
\beq
h_{c_2} = \epsilon^{\la_0+1}(k_u^{\la_0+1}) - \epsilon^{\la_0'}(k_u^{\la_0'})
+ {\cal U}(\la_0+1) + {\cal O}\left(U^2\right) \, .
\label{hc2genP}
\eeq
Eqs.\ (\ref{hc1genP}) and (\ref{hc2genP}) imply that
\beq
h_{c_2}-h_{c_1}
  = \epsilon^{\la_0+1}(k_u^{\la_0+1})-\epsilon^{\la_0}(k_u^{\la_0})
    + U O_1 +{\cal O}(U^2)
\eeq
with
\beq
O_1 = {1 \over 2 \pi p^2} \sum_{j=1}^p
   \left(\abs{a_{k_u^{\la_0+1},j}^{\la_0+1}}^2
     -   \abs{a_{k_u^{\la_0},j}^{\la_0}}^2 \right)
\left(
  \int_{k_l^{\la_0'}}^{k_u^{\la_0'}} d k' \abs{a_{k',j}^{\la_0'}}^2
   + \sum_{\la' < \la_0'} \int_{-\pi}^{\pi} d k' \abs{a_{k',j}^{\la'}}^2
  \right) \, .
\label{defO1}
\eeq
The first-order contribution does in general not vanish (for $p>2$).
It can be estimated as
\beq
\abs{O_1} \le {1 \over 2 \pi p^2} \sum_{j=1}^p p
 \left(
  \int_{k_l^{\la_0'}}^{k_u^{\la_0'}} d k' \abs{a_{k',j}^{\la_0'}}^2
   + \sum_{\la' < \la_0'} \int_{-\pi}^{\pi} d k' \abs{a_{k',j}^{\la'}}^2
  \right)
 = {n_{\dn} \over p} \, ,
\label{estO1}
\eeq
which shows that in principle it can be of order one.

\begin{figure}[hbt]
\centerline{\psfig{figure=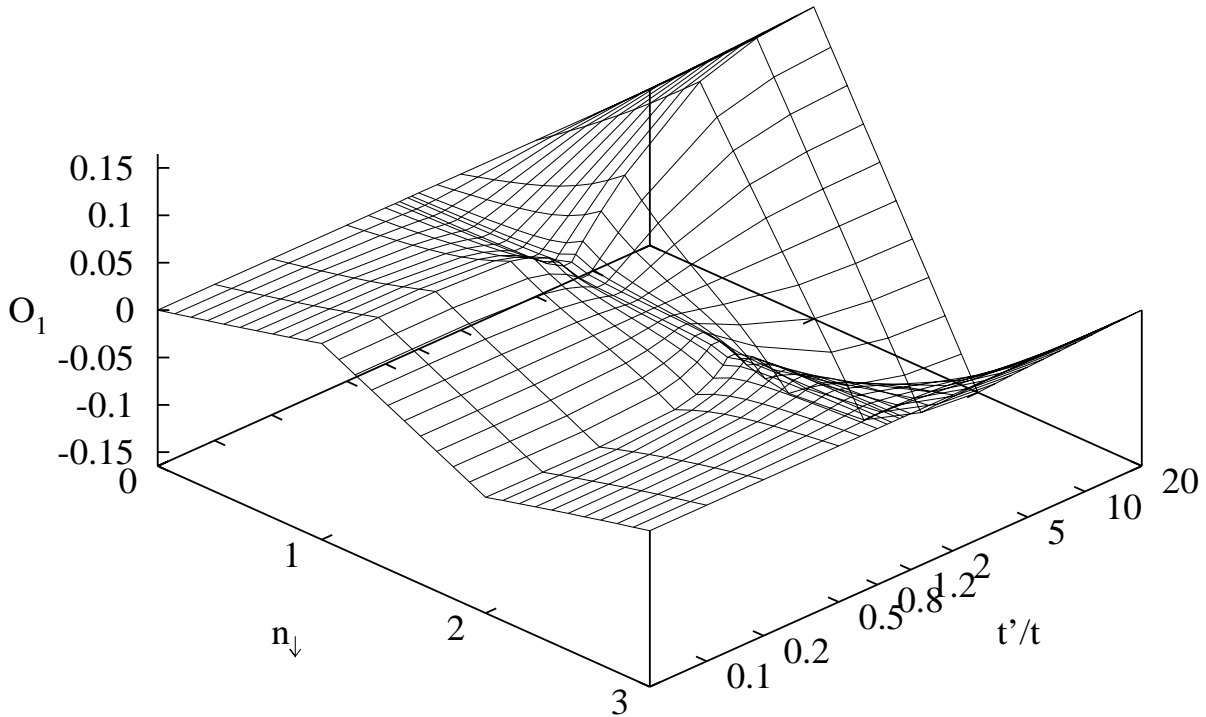,width=\columnwidth,angle=270}}
\smallskip
\caption{
Value of the first-order correction $O_1$ to the plateau width
as given by (\ref{defO1}) for $p=3$ and $n_\up = 2$.
\label{figO1}
}
\end{figure}

In general, it is not difficult to evaluate the first-order contribution
(\ref{defO1}) to the plateau width numerically. We will illustrate this now for
$p=3$. First notice that, for the case $p=3$,
the contribution from the kinetic energy can be readily evaluated as
\beq
\epsilon^3(0) - \epsilon^2(0) = \epsilon^2(\pi) - \epsilon^1(\pi) =
\abs{{\sqrt{8t^2+t'^2}-3 t' \over 2}} \, .
\eeq
Now we fix $n_\up = 2$, \ie\ the lowest two bands of up electrons are
completely filled. Then one has that $k_u^{3} = k_u^{2} = 0$ in
(\ref{defO1}). Numerical diagonalization of (\ref{su4}) and evaluation
of the remaining integrals in (\ref{defO1}) then leads to Fig.\
\ref{figO1}. Note that in the conventions of the other sections
(where $0 \le n \le 2$) this corresponds to the plateau with $m=4/3-n$.
The numerical data satisfies
$O_1 \to -O_1$ as $n_\dn \to 3-n_\dn$. This implies in particular
that the values of $O_1$ can be both positive and negative, corresponding to
an enhancement or reduction of the plateau width, respectively.
Furthermore, for $t \to 0$ and $n_\dn < 1$, the linear behaviour of
(\ref{estO1}) is reproduced, though with a coefficient which is
$1/6$, \ie\ by a factor of two smaller than in the estimate. The maximal
values attained are $\pm 1/6$ for $n_\dn \to 1$ or $2$, respectively
and $t \to 0$. This shows that the doping-dependent plateaux should
be stable features for $p=3$ as well.

We conclude this section by noting that
the calculations are also valid for the on-site $p$-merized
energy. The free Hamiltonian $H_0$ to be diagonalized is modified, but the
conclusions remain qualitatively unchanged.

\section{Lanczos diagonalization}
\label{secLanc}

Finally, we have performed Lanczos diagonalizations of the Hamiltonian
(\ref{pHub}) for $p=2$ and $p=3$ with constant $\mu(x) = \mu$ and
periodic boundary conditions on
finite lattices in order to further support the previous results. The
particle numbers $n_\up$ and $n_\dn$ have been used as quantum numbers
and translational symmetry was exploited. Furthermore, reflection symmetry
was exploited for $k=0,\pi$ and spin inversion for $n_\up = n_\dn$.

\subsection{Groundstate phase diagrams}
\label{secLancGS}

Computations have been performed mainly for one choice of parameters
due to the large number of sectors for which the groundstate energy
had to be found (for $p=2$ and $L=16$ of the order of $10^3$ sectors).
Keeping (\ref{JtU}) in mind,
we have chosen the parameters $U=3t$ and $\tp = 0.7t$ in order to look
at a situation sufficiently different from the limiting cases discussed
before, \ie\ both intermediate (to strong, as compared to the band-width)
on-site repulsion $U$ and intermediate $\tp/t$.

For the interpretation of our Lanczos results to be presented below, it
is useful to remember the following consequences of particle-hole symmetry on
a finite size lattice (see \cite{LW,Andrei} and references therein):
For $L$ even, the groundstate phase-diagram of the Hubbard chain with
periodic boundary conditions is symmetric under $\mu \mapsto
-U - \mu$ (with our conventions), while for $L$ odd the
particle-hole transformation interchanges
periodic and antiperiodic boundary conditions.

\begin{figure}[hbtp]
\centerline{\psfig{figure=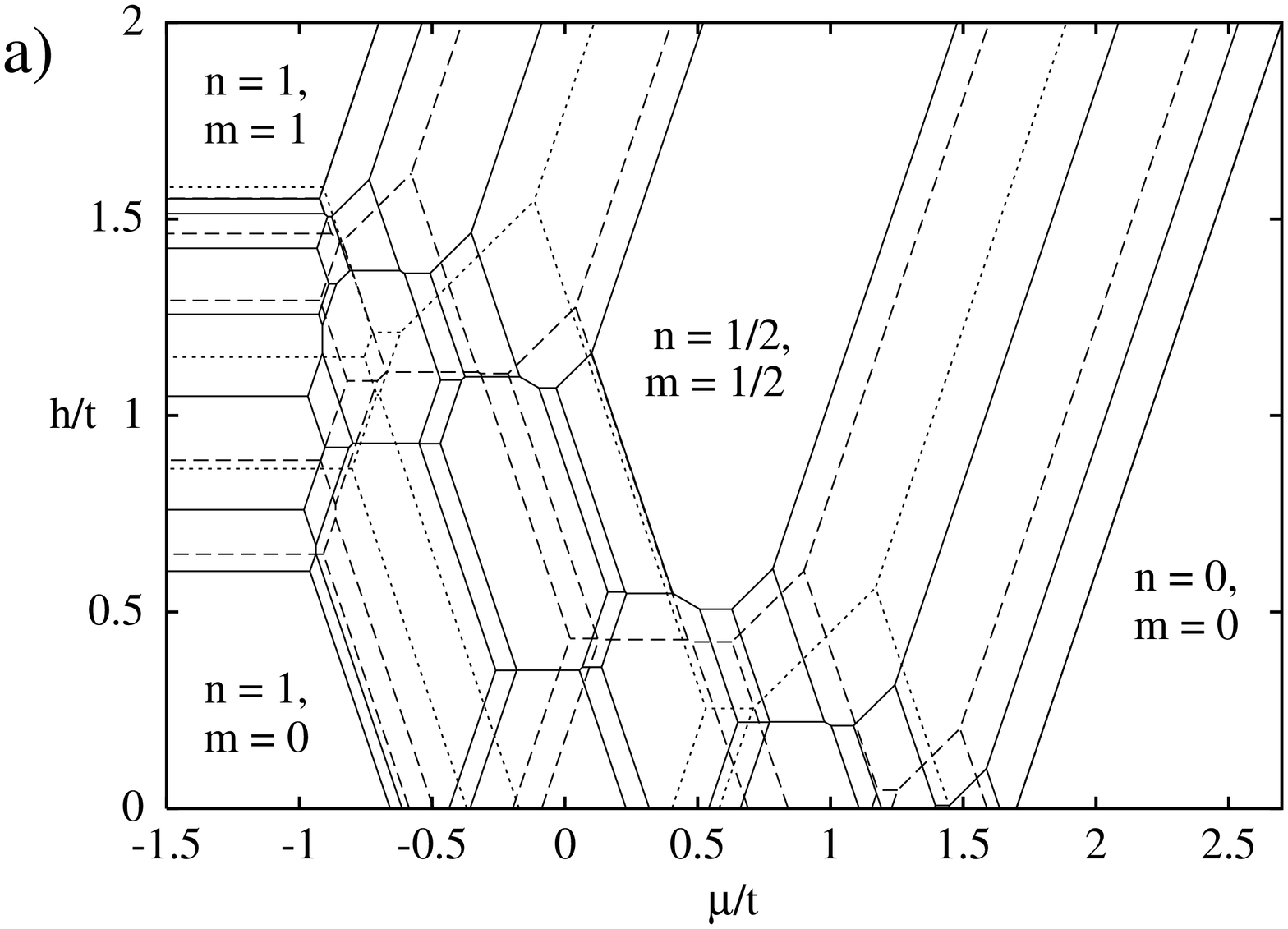,width=0.9\columnwidth,angle=270}}
\centerline{\psfig{figure=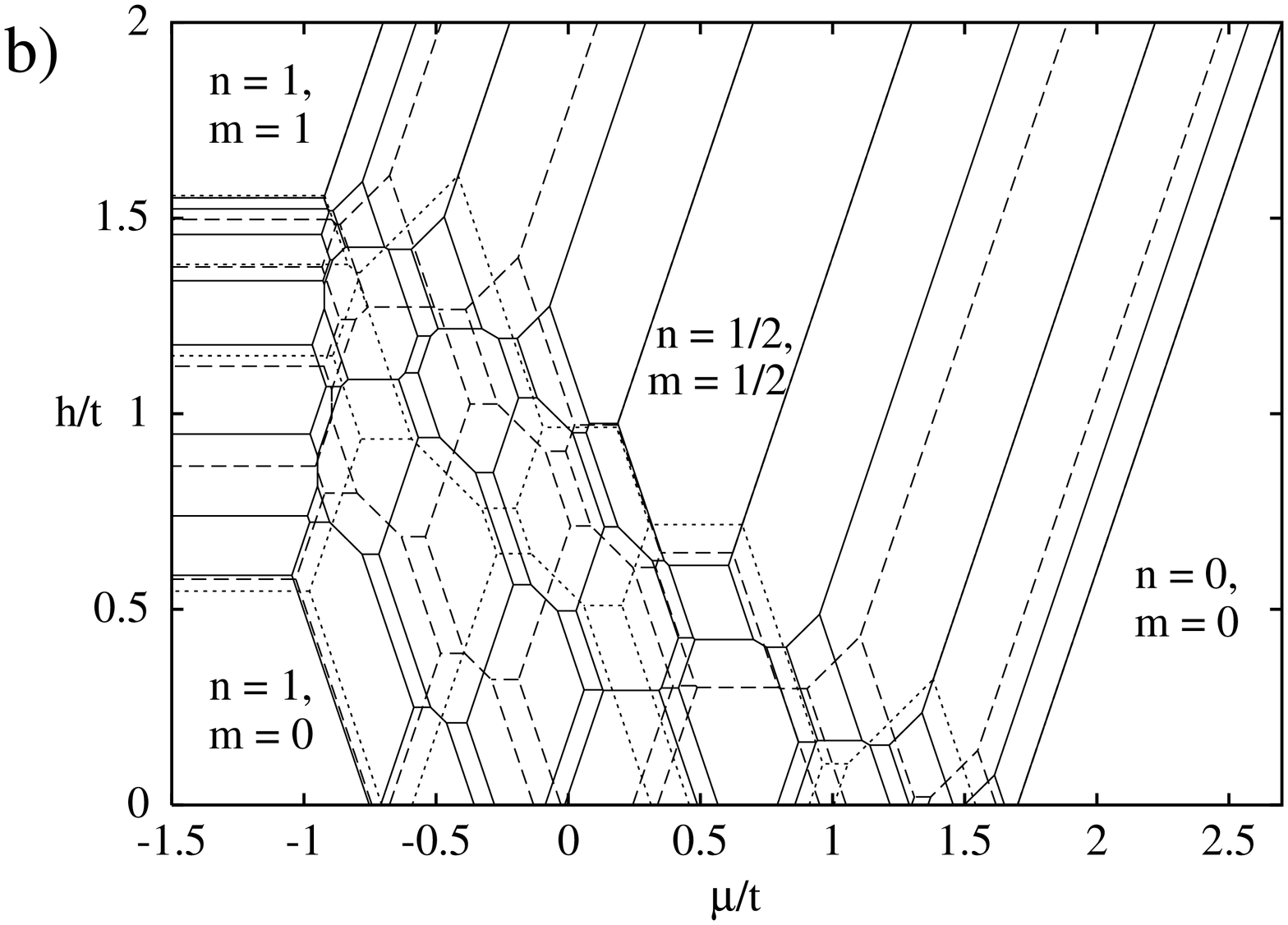,width=0.9\columnwidth,angle=270}}
\smallskip
\caption{
Groundstate phase diagram of the dimerized chain ($p=2$) with
$U=3t$, $\tp = 0.7t$. In a) the lines are for $L=6$ (dotted), $L=10$
(dashed) and $L=14$ (full) while in b) they are for $L=8$ (dotted), $L=12$
(dashed) and $L=16$ (full).
\label{p2gs}
}
\end{figure}

\begin{figure}[hbtp]
\centerline{\psfig{figure=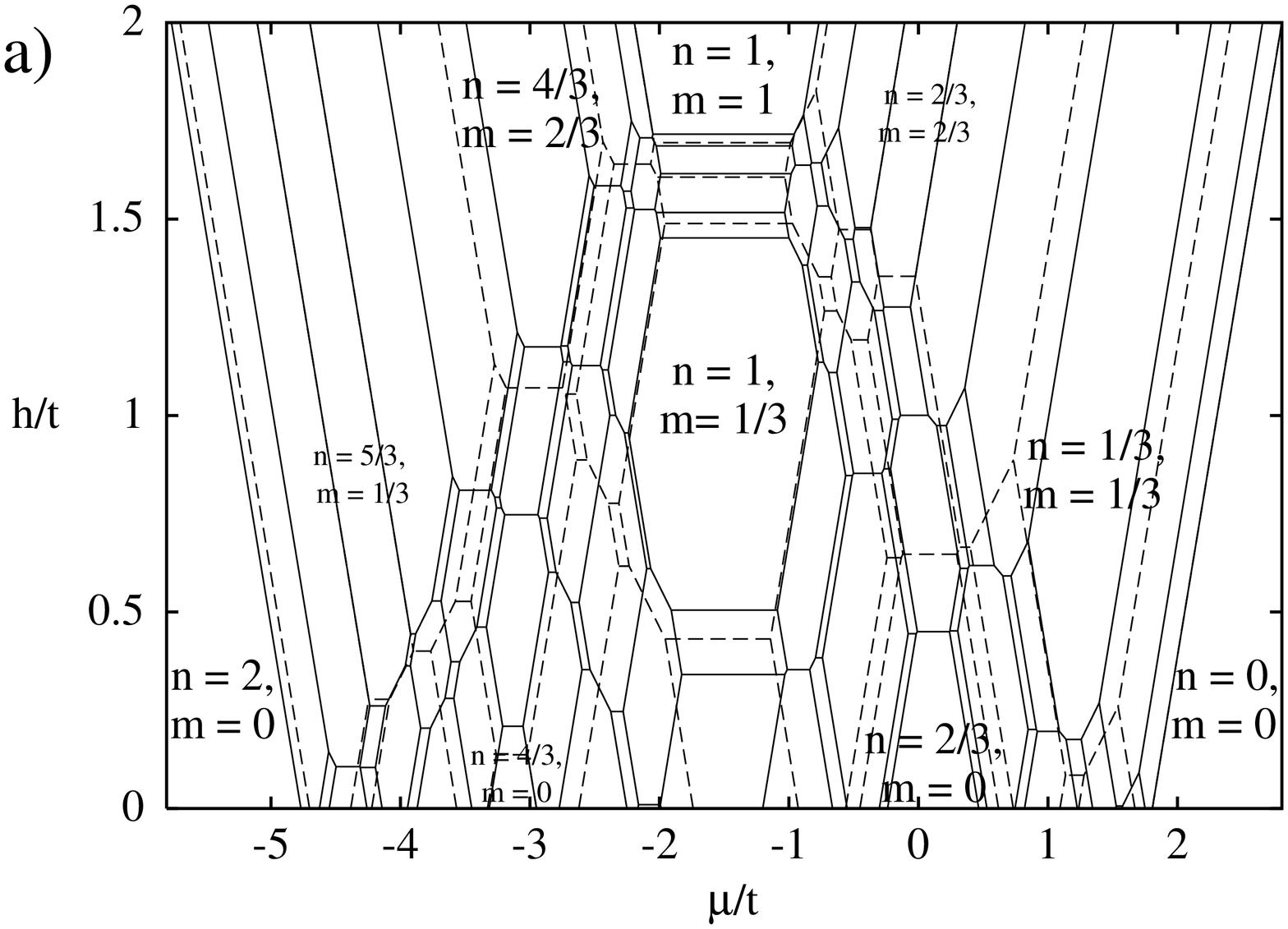,width=0.9\columnwidth,angle=270}}
\centerline{\psfig{figure=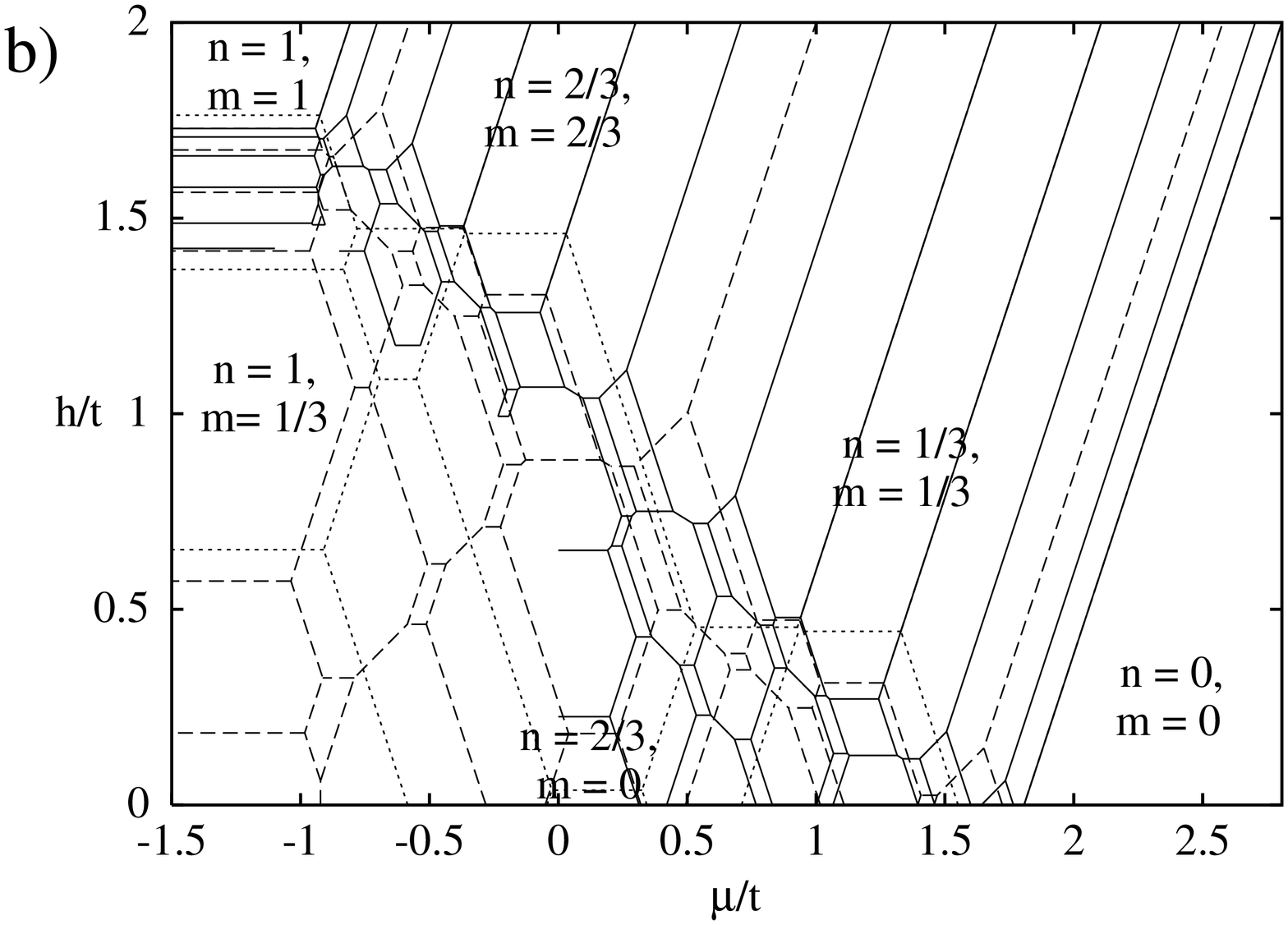,width=0.9\columnwidth,angle=270}}
\smallskip
\caption{
Groundstate phase diagram of the trimerized chain ($p=3$) with
$U=3t$, $\tp = 0.7t$. In a) the lines are for $L=9$ (dashed) and $L=15$ (full)
while in b) they are for $L=6$ (dotted), $L=12$ (dashed) and $L=18$ (full).
Note that the $L=18$ data in b) is incomplete for $n > 2/3$
({\it e.g.}\ only $m \ge 1/3$ for $n=1$).
\label{p3gs}
}
\end{figure}

Our numerical results for the groundstate phase diagram are
shown in Figs.\ \ref{p2gs} and \ref{p3gs} for $p=2$ and $p=3$, respectively.
The polygons in the figures denote regions in the ($\mu$,$h$)-plane
where the groundstate has a fixed filling $\nc$ and magnetization $m$ at
the given system size $L$ (those values of $m$ and $\nc$ which are common to
all investigated system sizes  are indicated in the figures). The groundstate
phase diagrams are symmetric under spin inversion ($m \to -m$ when $h\to -h$)
and as mentioned before, for even $L$ also under a particle-hole transformation
($\nc \to 2-\nc$ when $\mu \to -U-\mu$).
Therefore, for even $L$ we show only the quadrant with
$\mu \ge -U/2$, $h \ge 0$ and for odd $L$ only the region with $h \ge 0$.

The schematic groundstate phase diagrams in \cite{letter} were in fact
based on parts of these results and the reader may wish to use them
as a guide to the diagrams at finite size.

We note that for the saturated case $n_\dn = 0$ (and by particle-hole
symmetry also for $n_\up = 1$), the Coulomb repulsion is not effective
and the non-interacting result ($\epsilon^\pm(k)$ given by eq.\ (\ref{su6})
for $p=2$) can be used to determine the transitions between different
particle numbers. Complete agreement between this analytical computation
and the corresponding numerical results in Figs.\ \ref{p2gs} and \ref{p3gs}
is found. This also guides the interpretation of the finite-size data
since it follows in particular that the completely gapped
situations at saturation are those with $p \nc \in \Zed$ in the
thermodynamic limit. Such a guide is useful since the fermions behave
differently for even and odd particle numbers thus yielding
non-monotonic finite-size effects which can be still strong for the small
systems sizes considered here. In the particular case of the gapped states
with $p \nc \in \Zed$ at $m=\nc$ (or $m=2-\nc$), the corresponding
groundstate always has an even number of fermions when $L/p$ is even while
cases with an odd number occur when
$L/p$ is odd. This leads to vanishing finite-size effects for the
transition lines in the former case, but not always in the latter. Since even
and odd $L/p$ behave differently, we show separate figures for the
two cases.

For $p=2$, one can quite clearly recognize the fully gapped situations at
$(\nc,m) = (1,0)$, $(1,1)$, $(1/2,1/2)$ and $(0,0)$ from
the finite-$L$ data shown in Fig.\ \ref{p2gs}. Also the charge gap
at half filling ($\nc = 1$) is obvious. The most interesting region
is the doping-dependent plateau with $m=1-\nc$ which is a stable feature
in Fig.\ \ref{p2gs}a), but less clear in Fig.\ \ref{p2gs}b). Still, in
the latter case the region of stability of states with $m=1-\nc$ can
be seen to increase with increasing system size, thus supporting the
presence of a gap. Just the charge gap at quarter filling ($\nc = 1/2$)
is not distinct in this numerical data, but
it is known to be small anyway for these parameters \cite{PM}.

The case $p=3$ is shown in Fig.\ \ref{p3gs}. There is clear evidence for
the expected fully gapped situations (the labeled regions
in the figure) as well as the charge gap at half filling.
There is also evidence for the charge gap at $\nc = 2/3$ and the
equivalent case $\nc = 4/3$, just the charge gap at $\nc = 1/3, 5/3$
is again difficult to see. Also the expected doping-dependent
magnetization plateau with $m = \abs{\nc-2/3}$ can be recognized
in Fig.\ \ref{p3gs}a). By particle-hole symmetry, the plateau
with $m = \abs{4/3-\nc}$ must be present as well though it is more
difficult to recognize. The finite-size behavior of its
stability region in Fig.\ \ref{p3gs}a) (and of both plateaux in
the case of Fig.\ \ref{p3gs}b)) can again be taken as an indication
that it will indeed be present in the thermodynamic limit.

\subsection{Correlation functions}
\label{secLancCOR}

Having provided also numerical evidence for the existence of
doping-dependent magnetization plateaux, we now present a few numerical
results for correlation functions at $p=2$.

Is is technically useful to consider only objects which respect the
decomposition of the Hilbert space according to symmetries of the
Hamiltonian.  We therefore define averages of an
operator $A_x$ as
\beq
\left\langle A_x \right\rangle = \cases{
{1 \over L} \left\langle \psi_0(k) \right\vert \sum_{x_0 = 1}^L
  A_{x_0} \left\vert \psi_0(k) \right\rangle & for $k\ne 0,\pi$, \cr
{1 \over 2 L} \left\langle \psi_0(k) \right\vert \sum_{x_0 = 1}^L
  \left(A_{x_0} + A_{-x_0}\right)
 \left\vert \psi_0(k) \right\rangle & for $k=0,\pi$, \cr
}
\label{defAv}
\eeq
where $\left\vert \psi_0(k) \right\rangle$ is the groundstate
with momentum $k$. An additional advantage of this definition
is that oscillations originating from the modulation of $t(x)$
are smoothed by taken averages of the up to $p=2$ correlation
functions at a given distance.

The connected correlation function of two quantities $A$ and $B$ is defined as
\beq
C_{A,B}(x) = \left\langle A^\dagger_{x_0+x} B_{x_0} \right\rangle
         -\left\langle A^\dagger_{x_0}\right\rangle
          \left\langle B_{x_0} \right\rangle \, .
\label{defConCF}
\eeq
Of particular interest are the diagonal components of the superconducting
order parameter (\ref{defSC}) since quasi-long-range order is expected for
one of them.

\begin{figure}[hbt]
\centerline{\psfig{figure=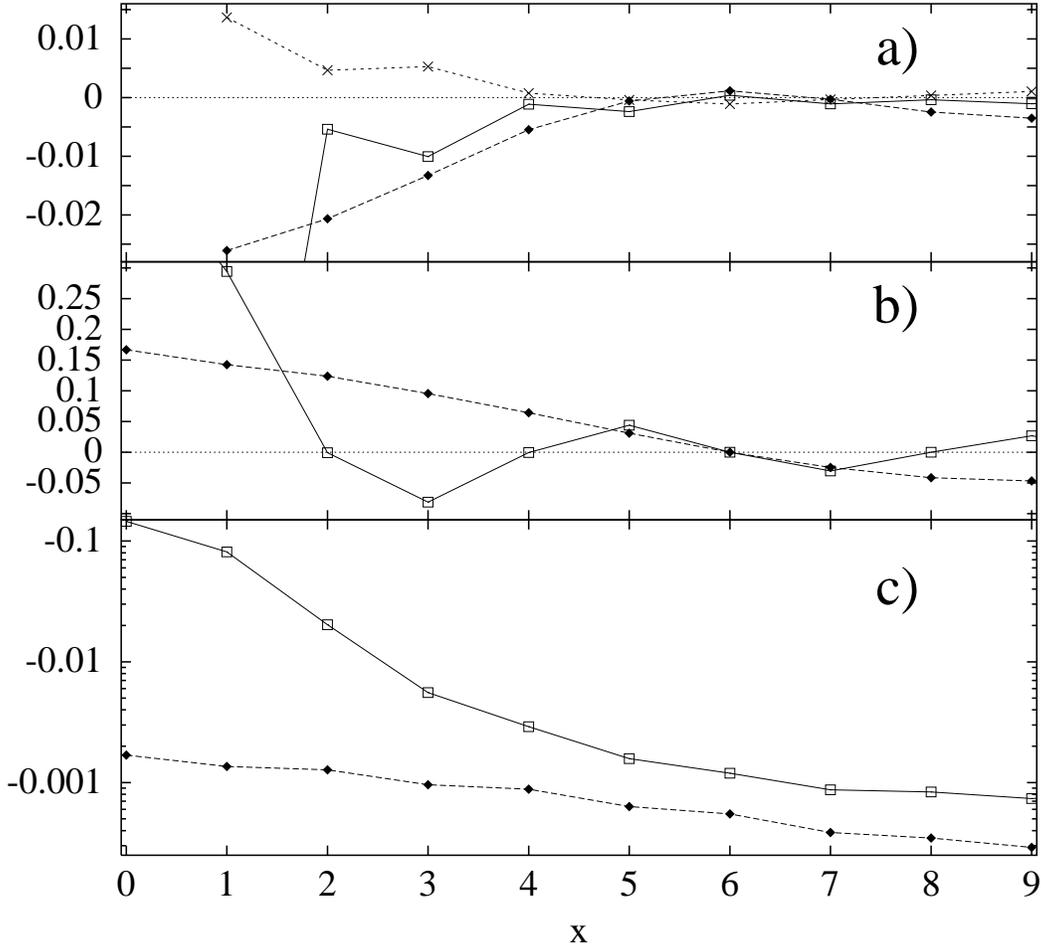,width=0.9\columnwidth,angle=270}}
\smallskip
\caption{
Correlation functions for the dimerized chain ($p=2$) with
$U=3t$, $\tp = 0.7t$ at $L=18$, $n=2/3$ and $m=1/3$.
Panel a) shows density-density correlations $C_{n_\alpha,n_\beta}(x)$,
panel b) electron-electron correlations $C_{c_\alpha,c_\beta}(x)$
and panel c) superconducting correlations $C_{\Delta_{\alpha,\alpha},
\Delta_{\beta,\beta}}(x)$. The symbols are for $\alpha=\beta=\up$
(boxes), $\alpha=\beta=\dn$ (filled diamonds) and $\alpha=\up$, $\beta=\dn$
(`$\times$'); the lines are guides to the eye.
\label{figCorrel}
}
\end{figure}

Numerical results for correlation functions on an $L=18$ system at the
plateau with $m=1-\nc$ are shown in Fig.\ \ref{figCorrel}. Characteristic
oscillations are observed in the density-density and electron-electron
correlation functions. This and the finite system size make a detailed
analysis of the asymptotics difficult. Nevertheless, one observes that
correlation functions containing up electrons decay faster than the
corresponding ones containing only down electrons (the latter may
still be smaller in absolute value due to a smaller overall prefactor).
In fact, all correlation functions shown in Fig.\ \ref{figCorrel} are very
similar to those obtained in the non-interacting situation ($U=0$) at the
same $L$. We therefore interpret our results as support for exponential
decay of all correlation functions containing $\up$ operators and
power-law decay for those containing only $\dn$ operators, as is
expected according to the analysis of section \ref{secAbbos}.
In particular, these numerical results are compatible with quasi-long-range
order in $\Delta_{\dn,\dn}$ on the $m=1-\nc$ plateau.

\section{Discussion and conclusions}
\label{secConcl}

We have shown that Hubbard chains with periodic hopping or on-site
energy present a rich structure of magnetization plateaux. More
precisely, for a periodicity $p$, we obtain the conditions
(\ref{cond}) for the appearance of the plateaux. If both
conditions are simultaneously satisfied, both spin and charge
degrees of freedom are massive. When the combination $p n \in
\Zed$ of these two conditions is satisfied, a charge gap opens
irrespective of the value of $m$. Finally, if just one of the
conditions (\ref{cond}) is satisfied, a magnetization plateau
appears {\it if} the total filling $\nc$ is kept fixed. This
result has been shown first by means of bosonization techniques,
valid in the regime where the differences in the modulation
amplitudes $\delta$ in (\ref{perturb}) and $\delta \mu$ in
(\ref{pHub1}) are small and for arbitrary values of $U$. We have
then shown that these results are confirmed by standard quantum
mechanical arguments valid for small $U$ and arbitrary
$p$-merization strength and provided an expression for the width
of the plateau to first order in $U$. We finally showed explicitly
such plateaux in finite size systems by means of Lanczos
diagonalization.

The combination of a gap which can be attributed to the spin degrees
of freedom and gapless (charge) modes prompted us to look
for superconducting correlations. Indeed, we found quasi-long-range order
in one component of the superconducting order parameter
in those cases where only one of the conditions
(\ref{cond}) is satisfied. 
%This is intriguing since it
%suggests that application of a magnetic field might induce superconductivity
%in a metal close to a Mott instability, but further
%theoretical investigations are needed to see if field-induced superconductivity
%is a realistic scenario.

In \cite{letter} it has been pointed out that the fully gapped situations
can be most easily understood in the limit $\tp = 0$ (the same argument
applies also for $\delta \mu = \infty$). Then the chain
effectively decomposes into clusters of $p$ sites, whose magnetization curves
are obviously staircase-like. The charge gap at $p \nc \in \Zed$
can also be understood in the limit of $\tp \ll t$ ($\delta \mu \gg \mu$),
one just needs to generalize the mapping of a quarter-filled dimerized
chain to an effective half-filled homogenous chain \cite{PM} to $m \ne 0$
and commensurate filling at general $p$. Finally, this mapping can also
be adapted to provide a further complementary argument for the existence of the
doping-dependent magnetization plateaux. Again, to first order in $\tp$,
an effective Hamiltonian can be found in the regime of strong $p$-merization,
\ie\ in the limit $ t' \ll t,U $ (resp. $ \delta \mu \gg t, U $)
for the case of modulation of the hopping amplitude (respectively of the
on-site energy). When only one of the conditions (\ref{cond}) is satisfied,
which amounts to a condition on the filling of spin-up or spin-down bands,
this effective Hamiltonian acquires a gap in the spin sector thus leading to a
doping-dependent magnetization plateau.

We would like to emphasize that such irrational plateaux are not present in
systems where the doping is not fixed. Moreover, due to the remaining massless
mode on such a plateau, the thermodynamical behavior of the system retains
some particularities of a gapless system, such as a specific heat vanishing
linearly as $T \to 0$.
An important feature is that the value of the magnetization $m$ on the plateaux
at fixed $\nc$ depends continuously on doping $\nc$. Analogous
situations \cite{FrSo,KLM} encourage us to believe that this scenario is
generic in doped systems. Doping could therefore be used as a tool
to study experimentally irrational plateaux in systems whose half-filled
parent compounds exhibit plateaux only at prohibitively high magnetic fields.
A natural candidate are ladders systems \cite{DR} where doping can indeed
be controlled. Theoretical results on doping-dependent magnetization plateaux
in Hubbard ladders will be reported elsewhere \cite{cdps}.

There are also natural problems for further study in the case of modulated
chains. For example, the large-$U$ limit of the Hubbard model leads to the
$t-J$ model. As a check of the generality of our results, one could therefore
investigate the $t-J$ model which at half filling would then be exactly
the situation studied in \cite{poly}. Due to the reduced Hilbert space,
the $t-J$ model would be particularly well suited for further numerical
checks. Another problem to be addressed is the universality class of the
transitions associated to the corners of a plateau. In the case of the
BA solvable model \cite{FrSo}, it was found that the presence of
a massless mode on a doping-dependent plateau may modify the universality
class of part of these transitions -- a fact which would be interesting to
investigate also in the present model.

\acknowledgements
We acknowledge useful discussions with P.\ Degiovanni, B.\ Dou\c{c}ot,
M.\ Fabrizio, E.\ Fradkin, H.\ Frahm, A.\ Izergin, F.\ Mila, T.M.\ Rice 
and C.\ Sobiella. This work has been done under partial support of the 
EC TMR Programme {\em Integrability, non-per\-turba\-tive effects and 
symmetry in Quantum Field Theories}.
A.H.\ is grateful to the Institut f\"ur Theo\-re\-tische Physik of the
ETH Z\"urich for hospitality and to the Alexander von Humboldt-foundation
for financial support during the initial stages of this project, as
well as to the C4 cluster of the ETH for allocation of CPU time.
The research of D.C.C. is partially supported by  CONICET,
ANPCyT (grant No.\ 03-02249) and Fundaci\'on Antorchas,
Argentina (grant No.\ A-13622/1-106).

\appendix

\section{Conventions}\label{convention}
In this short appendix, we define our conventions and notations.
The continuum fermion operators read
\bea
\psi_{\alpha} (x) = \ee^{-i k_{F,\alpha} x} \psi_{R,\alpha} (x) +
\ee^{i k_{F,\alpha} x} \psi_{L,\alpha} (x)  ~,\label{psiRL} \\
\psi_{\alpha}^\dagger (x) = \ee^{i k_{F,\alpha} x}
\psi_{R,\alpha}^{\dagger} (x) + \ee^{-i k_{F,\alpha} x}
\psi_{L,\alpha}^{\dagger} (x) ~.
\eea
Using standard bosonization rules we have:
\bea
\psi_{R,\alpha} (x) = \frac{1}{\sqrt{2\pi a} } \ee^{i \sqrt{4\pi}
\phi_{R,\alpha} (x)}+\ldots ~, \label{Rfer} \\
\psi_{L,\alpha} (x) = \frac{1}{\sqrt{2\pi a} } \ee^{-i \sqrt{4\pi}
\phi_{L,\alpha} (x)}+\ldots ~, \label{Lfer}
\eea
where $a$ is the lattice constant and $\phi_{R,L,\alpha}$ are the chiral
components of two real bosonic fields:
\beq
\phi_\alpha (x) = \phi_{R,\alpha} (x) + \phi_{L,\alpha} (x)~,
\eeq
whose dual fields are defined by
\beq
\theta_\alpha (x) = \phi_{R,\alpha} (x) - \phi_{L,\alpha} (x) ~.
\eeq
The dots in eqs.\ (\ref{Rfer}) and (\ref{Lfer}) stand for higher order terms,
due to the curvature of the dispersion relation, which are discussed in the
next appendix.
The up and down Fermi momenta are related to filling and magnetization:
\beq
k_+ = k_{F,\uparrow} + k_{F,\downarrow} = \pi n ~; \qquad
k_- = k_{F,\uparrow} - k_{F,\downarrow} = \pi m ~,
\eeq
where
\beq
n = \frac{1}{L} \langle \sum_{x,\alpha} n_{x,\alpha}  \rangle ~,
\qquad
m = \frac{2}{L} \langle \sum_{x} S^z_x \rangle
= {1 \over L} \langle \sum_{x,\alpha} n_{x,\up}-n_{x,\dn} \rangle ~,
\eeq
$n_{x,\alpha} = c^{\dagger}_{x,\alpha} c_{x,\alpha}$
and $L$ is the number of sites. Note that our definition of $m$
(which is the one used for the $XXZ$ chains in \cite{weSpin})
differs by a factor of $2$ from the one of Frahm and Korepin \cite{FK}.

\section{Fermion field operator}\label{boso}

In this appendix we discuss the bosonization of the fermion operator in the
Hubbard model in a magnetic field starting from the exact BA solution.
According to Frahm and Korepin \cite{FK}, the long-distance asymptotics of
zero-temperature correlation functions of physical fields is in general a sum
of terms of the form
\beq
\frac{\exp(- i 2 D_c
k_{F\uparrow} x) ~ \exp [- i ( D_c + D_s ) k_{F\downarrow} x] }
{(x - i v_c \tau)^{2 \Delta_c^+} (x + i v_c \tau)^{2 \Delta_c^-}
(x - i v_s \tau)^{2 \Delta_s^+} (x + i v_s \tau)^{2 \Delta_s^-} } ~,
\label{fk}
\eeq
where the scaling dimensions $\Delta^\pm_{c,s}$ are given by:
\bea
2\D_c^{\pm} & = & \left( Z_{cc}D_c+Z_{sc}D_s \pm
\frac{Z_{ss} \Delta N_c - Z_{cs} \Delta N_s}{2\det Z} \right)^2 + N_c^{\pm} ~, \\
2\D_s^{\pm} & = & \left( Z_{cs}D_c+Z_{ss}D_s \pm
\frac{Z_{cc} \Delta N_s - Z_{sc} \Delta N_c}{2\det Z} \right)^2 + N_s^{\pm} ~.
\eea
$\Delta N_{c,s}, D_{c,s}, N^\pm_{c,s}$ are the quantum numbers characterizing
the low energy excitations. $\Delta N_c$ and $\Delta N_s$ are integers
denoting the number of electrons and down spins with respect to the
groundstate and are fixed by the correlator
under consideration. The summation runs over all integers or half
integers $D_{c,s}$ (depending on the parity of $\D N_c,~\D N_s$) and
 on positive integers
$N_c^{\pm},~N_s^{\pm}$.

By analysing the leading contributions to the fermion two-point correlator,
one can write down, after some algebra, the bosonized fermion operator:
\bea
\psi_\downarrow = & \ee^{ - i k_{F \downarrow} x } ~ \ee^{ i
\sqrt{4 \pi} \phi_{R \downarrow}(x) } & ( r_1 + r_2 ~ \ee^{ - i 2 k_{F
\uparrow} x } ~ \ee^{ i \sqrt{4 \pi} \phi_\uparrow } + r_3 ~ \ee^{ i 2
k_{F \uparrow} x } ~ \ee^{ - i \sqrt{4 \pi} \phi_\uparrow } \nn \\
&& + r_4 ~ \ee^{ - i 2 k_{F \downarrow} x } ~ \ee^{  i \sqrt{4 \pi}
\phi_\downarrow } + r_5 ~ \ee^{ - i 2 k_+ x } ~ \ee^{  i \sqrt{4
\pi} (\phi_\uparrow + \phi_\downarrow) } + \ldots ) \nn \\
&+ \ee^{
i k_{F \downarrow} x } ~ \ee^{ - i \sqrt{4 \pi} \phi_{L
\downarrow}(x) } & ( l_1 + l_2 ~ \ee^{ i 2 k_{F \uparrow} x } ~ \ee^{ -
i \sqrt{4 \pi} \phi_\uparrow } + l_3 ~ \ee^{ - i 2 k_{F \uparrow} x
} ~ \ee^{  i \sqrt{4 \pi} \phi_\uparrow } \nn \\
&& + l_4 ~ \ee^{  i 2
k_{F \downarrow} x } ~ \ee^{ - i \sqrt{4 \pi} \phi_\downarrow } +
l_5 ~ \ee^{  i 2 k_+ x } ~ \ee^{ - i \sqrt{4 \pi} (\phi_\uparrow +
\phi_\downarrow) } + \ldots )
\label{renfer}
\eea
where $r_i, l_i$ are unknown numerical constants. Notice that at $U=0$,
$h=0$ all these constants vanish except $r_1=l_1= 1/\sqrt{2\pi
a}$. At $h=0$,
the scaling dimensions of the different contributions in eq.\ (\ref{renfer})
are known from BA for arbitrary repulsion $U$ and density $n \neq 1$.
It follows that it is sufficient to retain only the following terms
\bea
\psi_\downarrow &=& \ee^{ - i k_{F \downarrow} x } ~ \ee^{ i \sqrt{4 \pi}
\phi_{R \downarrow}(x) }
( r_1 + r_2 ~ \ee^{ - i 2 k_{F \uparrow} x } ~ \ee^{ i \sqrt{4 \pi} \phi_\uparrow }
+ r_3 ~ \ee^{ i 2 k_{F \uparrow} x } ~ \ee^{ - i \sqrt{4 \pi} \phi_\uparrow } +
 \ldots ) \nn \\
&& + \ee^{ i k_{F \downarrow} x } ~ \ee^{ - i \sqrt{4 \pi} \phi_{L \downarrow}(x) }
( l_1 + l_2 ~ \ee^{ i 2 k_{F \uparrow} x } ~ \ee^{ - i \sqrt{4 \pi} \phi_\uparrow }
+ l_3 ~ \ee^{ - i 2 k_{F \uparrow} x } ~ \ee^{  i \sqrt{4 \pi} \phi_\uparrow } +
\ldots )
\label{psid}
\eea
The expression for $\psi_{\uparrow}$ can be easily obtained by
exchanging $\downarrow$ and $\uparrow$, with the numerical constants
generically different.

Using this expression for $\psi_{\downarrow}$ and $\psi_{\uparrow}$,
one obtains:
\bea
\psi^\dagger_\downarrow \psi_{\downarrow} &=&
const ~ \partial_x \phi_{R\downarrow} + const ~ \partial_x \phi_{L\downarrow} +
2 r_1 l_1 ~ \sin ( 2 k_{F\downarrow} x - \sqrt{ 4 \pi } \phi_{\downarrow} )
\nn \\
&& + 2 ( r_1 l_2 + r_2 l_1) ~ \sin [ 2 k_+  x -  \sqrt{ 4 \pi }
( \phi_{\uparrow} + \phi_{\downarrow}) ]
\nn \\
&& - 2 ( r_1 l_3 + r_3 l_1) ~ \sin [ 2 k_-  x -  \sqrt{ 4 \pi }
( \phi_{\uparrow} - \phi_{\downarrow}) ] + \ldots
\label{psipsid}
\eea
and
\bea
\psi^\dagger_\uparrow \psi_{\uparrow} &=&
const ~ \partial_x \phi_{R\uparrow} + const ~ \partial_x \phi_{L\uparrow} +
2 r'_1 l'_1 ~ \sin ( 2 k_{F\uparrow} x - \sqrt{ 4 \pi } \phi_{\uparrow} )
\nn \\
&& + 2 ( r'_1 l'_2 + r'_2 l'_1) ~ \sin [ 2 k_+  x -  \sqrt{ 4 \pi }
( \phi_{\uparrow} + \phi_{\downarrow}) ]
\nn \\
&& + 2 ( r'_1 l'_3 + r'_3 l'_1) ~ \sin [ 2 k_-  x -  \sqrt{ 4 \pi }
( \phi_{\uparrow} - \phi_{\downarrow}) ] + \ldots
\label{psipsiu}
\eea
where we assumed the constants $r, l$ to be real. Otherwise,
the only modification would consist in shifts of the arguments of the
sines or cosines by unknown constant phases.

Now, assuming the constants to be equal for up and down fields,
and adding eqs.\ (\ref{psipsid}) and (\ref{psipsiu}) we obtain
for the bosonized density operator
\bea
\rho &=& \psi^\dagger_\uparrow \psi_{\uparrow}
      + \psi^\dagger_\downarrow \psi_{\downarrow} =
const ~ \partial_x ( \phi_\uparrow + \phi_\downarrow ) +
4 r_1 l_1 ~ \sin [ k_+ x- \sqrt{\pi}(\phi_\uparrow +\phi_\downarrow )]
 \nn  \\
&& \times \cos [ k_- x - \sqrt{\pi} ( \phi_\uparrow - \phi_\downarrow) ] +
4 ( r_1 l_2 + r_2 l_1)  ~ \sin [ 2 k_+ x - \sqrt{4\pi}
   ( \phi_\uparrow + \phi_\downarrow) ] + \ldots
\eea
Substituting finally (\ref{change1}), one obtains (\ref{density}).

Similarly, the difference of eqs.\ (\ref{psipsid}) and (\ref{psipsiu})
yields the $S^z$ operator:
\bea
2 S^z &=& \psi^\dagger_\uparrow \psi_{\uparrow} -
\psi^\dagger_\downarrow \psi_{\downarrow} = const ~ \partial_x (
\phi_\uparrow - \phi_\downarrow ) + 4 r_1 l_1 ~ \cos [ k_+ x-
\sqrt{\pi}(\phi_\uparrow +\phi_\downarrow )]
\nn \\
&& \times
\sin [ k_- x - \sqrt{\pi} ( \phi_\uparrow - \phi_\downarrow) ] - 4
( r_1 l_3 + r_3 l_1)  ~ \sin [ 2 k_- x - \sqrt{4\pi} (
\phi_\uparrow - \phi_\downarrow) ] + \ldots \label{sz}
\eea
Notice the last term in the $S^z$ operator. In the usual
treatments, (see for example \cite{GNT}), this term does not
appear. As it is obvious from eq.\ (\ref{sz}), this term would be
absent if we retain only the first two terms in eq.\ (\ref{psid}) or
if $r_1=l_1$ and $r_3 = - l_3$.

The assumptions on the constants $ r_i, l_i, r'_i,l'_i$ to be real
and equal for up and down fields are supported by %OPE
operator product expansion
computations of the original free fermion operator
with the perturbing Umklapp operator of the Hubbard Hamiltonian,
\beq
 \cos [2 k_+ x-
\sqrt{4\pi}(\phi_\uparrow +\phi_\downarrow )] \ .
\label{umklapp}
\eeq
These computations also show that, at lowest order, the
constants $r_2, r_3, l_2, l_3$ are linear in $U$.

\newpage
% A useful Journal macro
\def\Jrnl#1#2#3#4{{#1}{\bf #2}, #3 (#4)}
%
% Some useful journal names
%
\def\PLA{Phys.\ Lett.\ {\bf A}}
\def\PLB{Phys.\ Lett.\ {\bf B}}
\def\PR{Phys.\ Rev.}
\def\PRA{Phys.\ Rev. {\bf A}}
\def\PRB{Phys.\ Rev.\ {\bf B}}
\def\PRD{Phys. Rev. {\bf D}}
\def\PRL{Phys.\ Rev.\ Lett.\ }
\def\EPJB{Eur.\ Phys.\ J.\ {\bf B}}
\def\RMP{Rev.\ Mod.\ Phys.\ }
\def\JPSJ{J.\ Phys.\ Soc.\ Jpn.\ }
\def\NPB{Nucl.\ Phys.\ {\bf B}}


\begin{thebibliography}{99}


\bibitem[\ast]{PAADM} Present address: 
Fakult\"at f\"ur Physik, Albert-Ludwigs-Universit\"at, 
79104 Freiburg, Germany.

\bibitem[\dagger]{PAPS} Present address: Department of Physics and Astronomy, 
University of British Columbia, V6T 1Z1 Vancouver, B.C., Canada.

\bibitem[\ddagger]{URA} URA 1325 du CNRS associ\'ee \`a l'Ecole Normale 
Sup\'erieure de Lyon.

\bibitem{HiOk} K.\ Hida, 
{\Jrnl {\JPSJ} {63} {2359} {1994}}; 
K.\ Okamoto, 
Solid State Commun.\ {\bf 98}, 245 (1996).

\bibitem{AOY} M.\ Oshikawa, M.\ Yamanaka and I.\ Affleck,
{\Jrnl {\PRL} {78} {1984} {1997}}.

\bibitem{Totsuka} K.\ Totsuka, 
{\Jrnl {\PLA} {228} {103} {1997}};
{\Jrnl {\PRB} {57} {3454} {1998}};
{\Jrnl {\EPJB} {5} {705} {1998}}.

\bibitem{weSpin} D.C.\ Cabra, A.\ Honecker and P.\ Pujol,
{\Jrnl {\PRL} {79} {5126} {1997}};
{\Jrnl {\PRB} {58} {6241} {1998}};
{\Jrnl {\EPJB}  {13} {55} {2000}}.

\bibitem{FGKMW} A.\ Fledderjohann, C.\ Gerhardt, M.\ Karbach, 
K.-H.\ M\"utter, and R. Wie\ss ner,
{\Jrnl {\PRB} {59} {991} {1999}}.

\bibitem{poly} D.C.\ Cabra and M.D.\ Grynberg,
{\Jrnl {\PRB} {59} {119} {1999}}; 
A.\ Honecker,
{\Jrnl {\PRB} {59} {6790} {1999}}.

\bibitem{YaSa} S.\ Yamamoto and T.\ Sakai, 
{\Jrnl {\PRB} {62} {3795} {2000}}.

\bibitem{S1dim} Y.\ Narumi, M.\ Hagiwara, R.\ Sato, K.\ Kindo, 
H.\ Nakano and M.\ Takahashi,
Physica {\bf B246-247}, 509 (1998).

\bibitem{Ishii} M.\ Ishii, H.\ Tanaka, M.\ Hori, H.\ Uekusa, 
Y.\ Ohashi, K.\ Tatani, Y.\ Narumi and K.\ Kindo, 
{\Jrnl {\JPSJ} {69} {340} {2000}}.

\bibitem{fTri} K.\ Okamoto, T.\ Tonegawa, Y.\ Takahashi and M.\ Kaburagi 
J.\ Phys.: Condensed Matter {\bf 11}, 10485 (1999); 
T.\ Tonegawa, K.\ Okamoto, T.\ Hikihara, Y.\ Takahashi, 
M.\ Kaburagi, 
J.\ Phys.\ Soc.\ Jpn.\ {\bf 69} Suppl.\ A, 332 (2000);
%preprint cond-mat/9912482
K.\ Sano and K.\ Takano, 
{\Jrnl {\JPSJ} {69} {2710} {2000}};
%preprint cond-mat/0005281;
A.\ Honecker and A.\ L\"auchli, preprint cond-mat/0005398.

\bibitem{srcu2bo32} H.\ Kageyama, K.\ Yoshimura, R.\ Stern,
N.V.\ Mushnikov, K.\ Onizuka, M.\ Kato, K.\ Kosuge, C.P.\ Slichter, 
T.\ Goto and Y.\ Ueda, 
{\Jrnl {\PRL} {82} {3168} {1999}};
K.\ Onizuka, H.\ Kageyama, Y.\ Narumi, K.\ Kindo, Y.\ Ueda 
and T.\ Goto 
{\Jrnl {\JPSJ} {69} {1016} {2000}}.

\bibitem{STKT} W.\ Shiramura, K.\ Takatsu, B.\ Kurniawan, H.\ Tanaka, 
H.\ Uekusa, Y.\ Ohashi, K.\ Takizawa, H.\ Mitamura and T. Goto, 
{\Jrnl {\JPSJ} {67} {1548} {1998}}.

\bibitem{2Dtheory} S.\ Miyahara and K.\ Ueda,  
{\Jrnl {\PRL} {82} {3701} {1999}}; 
T.\ Momoi and K.\ Totsuka, 
{\Jrnl {\PRB} {61} {3231} {2000}}; 
E.\ M\"uller-Hartmann, R.R.P.\ Singh, C.\ Knetter and G.S.\ Uhrig, 
{\Jrnl {\PRL} {84} {1808} {2000}}; 
T.\ Momoi and K.\ Totsuka, {\Jrnl {\PRB} {62}{15067}{2000}}.
%preprint cond-mat/0006020.

\bibitem{DR} E.\ Dagotto and T.M.\ Rice,
{\Jrnl {Science } {271} {618} {1996}};
T.M.\ Rice, 
Z.\ Phys.\ {\bf B103}, 165 (1997).

\bibitem{letter} D.C.\ Cabra, A.\ De\ Martino, A.\ Honecker,
P.\ Pujol and P.\ Simon,
{\Jrnl {\PLA} {268} {418} {2000}}.

\bibitem{FrSo} H.\ Frahm and C.\ Sobiella,
{\Jrnl {\PRL} {83} {5579} {1999}}.

\bibitem{KLM} H.\ Tsunetsugu, M.\ Sigrist and K.\ Ueda,
{\Jrnl {\RMP} {69} {809} {1997}}.

\bibitem{orgS} J.T.\ Devreese, R.P.\ Evrard and V.E.\ van Doren (eds.),
{\it Highly Conducting One-Dimensional Solids}, Plenum Press, New York (1979);
T.\ Ishiguro and K.\ Yamaji, {\it Organic Superconductors},
Springer Series in Solid-State Sciences 88, Berlin (1990).

\bibitem{EIT} T.\ Egami, S.\ Ishihara and M.\ Tachiki,
{\Jrnl {Science } {261} {1307} {1993}}.

\bibitem{BWRZZHD} M.R.\ Bond, R.D.\ Willett, R.S.\ Rubins, P.\ Zhou, 
C.E.\ Zaspel, S.L.\ Hutton and J.E.\ Drumheller,
{\Jrnl {\PRB} {42} {10280} {1990}}.

\bibitem{AAIAG} Y.\ Ajiro, T.\ Asano, T.\ Inami, H.\ Aruga-Katori and
T.\ Goto, 
{\Jrnl {\JPSJ} {63} {859} {1994}}.

\bibitem{LW} E.\ Lieb and F.Y.\ Wu,
{\Jrnl {\PRL} {20} {1445} {1968}}.

\bibitem{FK} H.\ Frahm and V.E.\ Korepin,
{\Jrnl {\PRB} {42} {10553} {1990}};
{\Jrnl {\PRB} {43} {5653} {1991}}.

\bibitem{EF} F.H.L.\ Essler and H.\ Frahm,
{\Jrnl {\PRB} {60} {8540} {1999}}.

\bibitem{PS} K.\ Penc and J.\ S\'olyom, 
{\Jrnl {\PRB} {47} {6273} {1993}}.

\bibitem{GNT} A.O.\ Gogolin, A.A.\ Nersesyan and A.M.\ Tsvelik,
{\it Bo\-so\-ni\-za\-tion and Strongly Correlated Electron Systems},
Cambridge University Press, Cambridge (1998).

\bibitem{Schulz} H.J.\ Schulz, 
Int.\ J.\ Mod.\ Phys.\ {\bf B5}, 57 (1991); 
p.\ 533 in {\it Proceedings of Les Houches Summer School LXI}, 
eds.\ E.\ Akkermans, G.\ Montambaux, J.\ Pichard and J.\ Zinn-Justin 
(Elsevier, Amsterdam, 1995) [{\tt cond-mat/9503150}].

\bibitem{Voit} J.\ Voit, 
Rep.\ Prog.\ Phys.\ {\bf 58}, 977 (1995).

\bibitem{conv} $m$ is normalized to its saturation value.
For our conventions on filling and magnetization see appendix \ref{convention}.

\bibitem{PM} K.\ Penc and F.\ Mila,
{\Jrnl {\PRB} {50} {11429} {1994}}.

\bibitem{NO99} S.\ Nishimoto and Y.\ Ohta,
{\Jrnl {\PRB} {59} {4738} {1999}}.

\bibitem{DL99} M.L.\ Doublet and M.B.\ Lepetit,
{\Jrnl {J.\ Chem.\ Phys. } {110} {1767} {1999}}.

\bibitem{NTO99} S.\ Nishimoto, M.\ Takahashi and Y.\ Ohta,
{\Jrnl {\JPSJ} {69} {1594} {2000}}.

\bibitem{Comprad} The compactification radius $R$ and the
parameter $\xi$ in eq.\ (\ref{chargespin}) are related to the standard
Luttinger parameter $K$ respectively as $1/(2\pi R^2) = K$ and $\xi^2=2K$.

\bibitem{FGN} M.\ Fabrizio, A.O.\ Gogolin and A.A.\ Nersesyan,
{\Jrnl {\PRL} {83} {2014} {1999}}.

\bibitem{cdm} P.\ Degiovanni, Ch.\ Chaubet and  R.\ Melin,
{\Jrnl {Theor.\ Math.\ Phys.\ } {117} {5} {1998}}.

\bibitem{Tan} K.\ Tandon, S.\ Lal, S.K.\ Pati, S.\ Ramasesha, and D. Sen,
{\Jrnl {\PRB} {59} {396} {1999}}.

\bibitem{nanoT} D.\ Green and C.\ Chamon,% preprint cond-mat/0004292.
{\Jrnl {\PRL} {85}{4128}{2000}}.

\bibitem{Kagome} P.\ Lecheminant, B.\ Bernu, C.\ Lhuillier, 
L.\ Pierre and P.\ Sindzingre, 
{\Jrnl {\PRB} {56} {2521} {1997}}; 
C.\ Waldtmann, H.-U.\ Everts, B.\ Bernu, C.\ Lhuillier, 
P.\ Sindzingre, P.\ Lecheminant, L.\ Pierre:
{\Jrnl {\EPJB} {2} {501} {1998}}.

\bibitem{totnum} The total number of particles denoted by $N$ is given by
$N = \sum_{\la,\sigma} N_\sigma^\la$.

\bibitem{normden} In this section we use the notation $n_\sigma =
\sum_{\la} n_\sigma^\la$, \ie\ the normalization is such that $0
\le n_\sigma \le p$.

\bibitem{Andrei} N.\ Andrei, p.\ 1 in 
{\it Summer Course on Low-Di\-men\-sional Quantum Field Theories 
for Condensed Matter Physicists}, Trieste (1992), 
eds.\ S.\ Lundqvist, G.\ Morandi and Yu Lu 
(World Scientific, Singapore 1995) [{\tt cond-mat/9408101}].

\bibitem{cdps} D.C.\ Cabra, A.\ De\ Martino,
P.\ Pujol and P.\ Simon, preprint cond-mat/0012235.

\end{thebibliography}
\end{document}